%% file: main.tex
\documentclass{article}
\usepackage{PRIMEarxiv}
\usepackage[utf8]{inputenc}
\usepackage[T1]{fontenc}
\usepackage{amsmath,amssymb,booktabs,array,tabularx}
\usepackage{graphicx,microtype}
\usepackage[numbers,square]{natbib}
\usepackage{placeins}
\usepackage[hidelinks]{hyperref}
\newcommand{\tmodel}{text2midi}
\newcommand{\mmodel}{MIDI-LLM}
\newcommand{\RCE}{R_{\mathrm{CE}}}
\newcommand{\inputrows}[1]{\csname @@input\endcsname #1 }
\title{MI-MIDI 2: From Observation to Control\\
with Sparse Autoencoders in Text-to-MIDI Generation}
\author{
  Jakub Poćwiardowski, Mateusz Modrzejewski \\
  Institute of Computer Science \\
  Warsaw University of Technology \\
  Warsaw, Poland \\
  \texttt{j.pocwiardo@gmail.com, mateusz.modrzejewski@pw.edu.pl} \\
}
\date{}

\begin{document}
\maketitle
\begin{abstract}
Sparse autoencoders (SAEs) can expose features of a generative model without
predefined labels, but in language models the evidence used to interpret these
features has proven an unreliable guide to which of them steer the output.
Whether interpretations of SAE features predict control in music generators
has not, to our knowledge, been tested. We develop an end-to-end protocol for
discovering and evaluating SAE features in text-to-MIDI generation and apply it
to \tmodel{} and \mmodel{}. The protocol characterizes features through
correlations with musical metrics, shared instrument programs, and automated
descriptions, each with its own control: note-density adjustment, matched
inactive excerpts, and permutation calibration. It then tests candidates by
intervention, measured directly in the generated MIDI against unmodified
generation and a matched random direction, ranks them in screens, and confirms
selected features on new prompts. Evidence that survives these controls points
to features that steer the output. Descriptions exceed permutation controls
for instrumentation, and a feature whose description names a synth-lead
program introduces that program in a prespecified confirmation and in a
replication on disjoint prompts and reappears in independently trained
dictionaries. Features selected by density-adjusted correlations show a small
positive shift of their target metric as a family, whereas features selected
by raw correlations show none. An atlas provides MIDI, piano-roll, and audio
examples for every feature of the targeted screen.
\end{abstract}


\input{sections/introduction}
\input{sections/related}
\input{sections/setup}
\input{sections/observation}
\input{sections/interventions}
\input{sections/discussion}

\section*{Acknowledgments}
We gratefully acknowledge Polish high-performance computing infrastructure PLGrid (HPC Centers: ACK Cyfronet AGH) for providing computer facilities and support within computational grant no. PLG/2025/018892.

\bibliographystyle{unsrtnat}
\bibliography{references}
\clearpage
\appendix
\input{sections/appendix}
\end{document}

%% file: sections/introduction.tex
\section{Introduction}
\label{sec:intro}

Interpretability methods offer a route to controlling text-conditioned music
generators. Adding a hidden-state direction associated with a musical attribute
changes that attribute during generation, as shown for instrumentation, genre,
tempo, and timbral brightness in text-to-audio models
\citep{vasquez2024exploring,facchiano2025activation}; such interventions have
also been evaluated in text-to-MIDI models \citep{pocwiardowski2026mimidi}. These
directions are supervised: the attribute must be chosen in advance. Sparse
autoencoders (SAEs) remove this requirement. They decompose hidden states into
a large dictionary of sparsely active directions learned without labels
\citep{bricken2023monosemanticity,huben2024sparse}, and each dictionary
direction can itself steer the model \citep{templeton2024scaling}.

To use a dictionary feature for control, one must know what it represents and
whether that interpretation predicts its effect. The evidence for a feature's
meaning is mostly observational: the examples on which it is most active, its
correlation with a measurable property, or a description that a language model
writes from those examples \citep{bills2023language,paulo2025automatically}.
In language models, this evidence is an unreliable guide to control. Features
that respond strongly to inputs differ from those that steer outputs
\citep{arad2025saes}, descriptions based on inputs often miss a feature's
effect on generation \citep{gurarieh2025enhancing}, SAE steering
underperforms simple baselines \citep{wu2025axbench}, and automated
interpretability scores do not separate trained from randomly initialized
transformers \citep{heap2026automated}.

For music generators, this relation has not been tested. SAE features of
MusicGen have been described by a multimodal language model and used for
steering, with the effect judged from the generated audio
\citep{singh2026discovering}, and SAEs of a symbolic generator have been used
to steer attributes defined in advance \citep{prokopiou2026closing}. To our
knowledge, no study has asked whether the observational evidence about
unsupervised features of a music generator predicts their effect on the
generated music. Text-to-MIDI generators make this question answerable,
because their output is a score whose notes, durations, and instrument
programs can be measured exactly.

\begin{figure}[t]
\centering
\includegraphics[width=\linewidth]{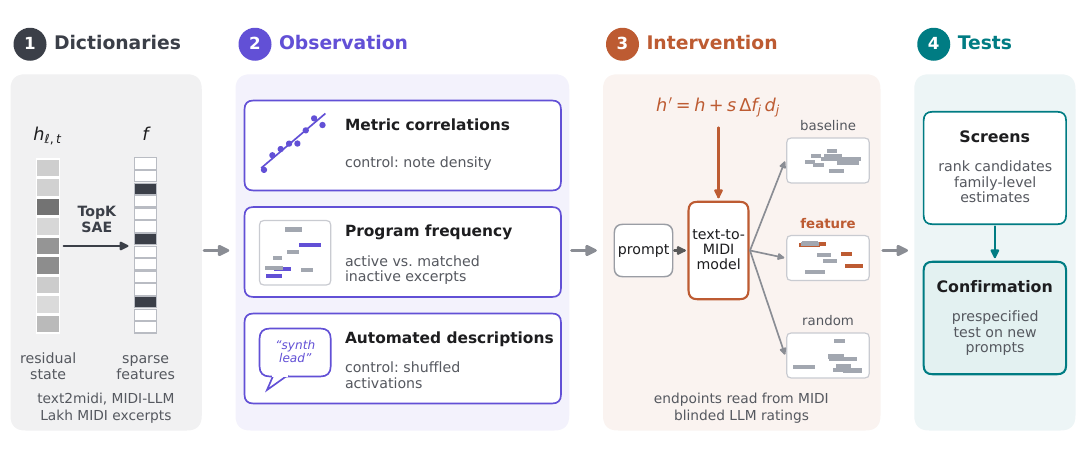}
\caption{Overview of the experiments. Each observational route has its own
control; interventions compare the feature direction with baseline generation
and a matched random direction, with endpoints measured in MIDI. Screens rank
candidates and give family-level estimates; selected features are confirmed on
new prompts.}
\label{fig:overview}
\end{figure}

We study SAE features of two text-to-MIDI models of contrasting design,
\tmodel{} \citep{bhandari2025text2midi} and \mmodel{} \citep{wu2026midi}, in
the four stages of Figure~\ref{fig:overview}. After training the dictionaries,
we collect three kinds of observational evidence about their features:
correlations with symbolic music metrics, instrument programs shared by a
feature's active excerpts, and automated descriptions tested on held-out
excerpts. Each kind needs a control. Many musical statistics can grow with the
number of notes in an excerpt, so we adjust correlations for note density, and
a language model may find a pattern in any group of excerpts, so we compare
descriptions with permutation controls. We then steer generation along the
features' decoder directions and measure the outcome in MIDI against
unmodified generation and a matched random direction. We screen many features
on few prompts to rank candidates and estimate effects for families of
features, and we confirm individual features on new prompts.

Our contributions are as follows.
\begin{enumerate}
    \item \textbf{A validation protocol for SAE features of symbolic music
    generators.} It pairs each kind of observational evidence with a control
    and measures interventions in the generated MIDI against two references,
    so that researchers can locate the step at which an interpretation stops
    predicting the music.
    \item \textbf{Evidence that controlled observations identify steerable
    features.} A feature from the only description group that survives the
    permutation control steers its instrument program in a prespecified
    confirmation and a replication, and features selected after the density
    adjustment show a small positive family-level shift of their target
    metric, unlike features selected by raw correlations. Practitioners can
    therefore select steering candidates by evidence that survives its control
    and confirm them on new prompts.
    \item \textbf{To our knowledge, the first SAE dictionaries and feature atlas
    for text-conditioned symbolic music generators.} The atlas
    (\url{https://jpocwiar.github.io/MI-MIDIv2-Demo/}) provides MIDI files,
    piano rolls, and audio for all 80 features of the targeted screen, so that
    readers can inspect the music behind the measurements.
\end{enumerate}

%% file: sections/related.tex
\section{Related Work}
\label{sec:related}

\paragraph{Sparse autoencoders.}
Neural networks can represent more features than they have dimensions, so
individual neurons often respond to several unrelated concepts
\citep{elhage2022toy}. SAEs address this superposition with an overcomplete
dictionary of sparse coefficients
\citep{bricken2023monosemanticity,huben2024sparse,templeton2024scaling}. We use
the TopK activation \citep{makhzani2013k} as adapted to language-model
activations by \citet{gao2025scaling}; JumpReLU is an alternative
\citep{rajamanoharan2024jumping}. Following \citet{gao2025scaling}, we
evaluate reconstruction error and the preservation of downstream loss
separately. Our held-out recognition task is a variant of detection scoring,
in which a language model describes a feature from activating examples and
the description is scored by how well it predicts activation on new examples
\citep{bills2023language,paulo2025automatically}. Because SAEs trained with
different seeds learn partly different features \citep{paulo2026sparse}, we
also check whether a confirmed feature reappears in independently trained
dictionaries.

\paragraph{Steering with dictionary features.}
Activation steering adds a direction to the hidden state during generation
\citep{turner2023steering}, commonly a contrastive or difference-in-means
direction \citep{rimsky2024steering}. For SAE decoder directions,
\citet{arad2025saes} select features by an output score and
\citet{gurarieh2025enhancing} describe features by their effect on outputs.
\citet{peng2026position} argue that SAEs are better suited to discovering
unanticipated concepts than to acting on known ones.

\paragraph{Music generators.}
Most mechanistic work on music generation studies MusicGen
\citep{copet2023simple}, including steering of instrumentation and genre
\citep{vasquez2024exploring}, of tempo and timbral brightness
\citep{facchiano2025activation}, and with SAE decoder directions
\citep{singh2026discovering}; \citet{chen2026dual} separate work identity from
performance variation with dual-contrastive SAEs. In symbolic generation,
\citet{prokopiou2026latent} steer pitch and duration in the Multitrack Music
Transformer with difference-in-means directions, and
\citet{prokopiou2026closing} train TopK SAEs on the same model and use feedback
control to steer predefined attributes. Both models we study build on the
captioned MIDI of MidiCaps \citep{melechovsky2024midicaps}. Our earlier MI-MIDI
study analyzes them with probing, logit and tuned lenses, and activation
patching, and evaluates difference-in-means steering in both orientations
\citep{pocwiardowski2026mimidi}.

%% file: sections/setup.tex
\section{Models and Dictionaries}
\label{sec:setup}

\paragraph{Models.}
\tmodel{} conditions an 18-layer decoder of residual width 768 on a frozen
Flan-T5 encoder and generates REMI+ tokens with explicit bar, position,
instrument, pitch, velocity, and duration events
\citep{bhandari2025text2midi,von2023figaro}. \mmodel{} is the 16-layer
Llama~3.2~1B decoder of residual width 2048 with the vocabulary of the
Anticipatory Music Transformer, which encodes absolute onset time, duration,
and joint instrument--pitch events \citep{wu2026midi,thickstun2023anticipatory}.
We analyze layers 2, 8, 13, and 16 of \tmodel{} and layers 2, 7, 12, and 14 of
\mmodel{}. Dictionaries, characterization, and interventions all use the
residual state $h_{\ell,t}$ at the output of layer $\ell$ after the model has
consumed MIDI token $t$.

\paragraph{Data.}
We use 1,500 files of the Lakh MIDI Dataset \citep{raffel2016lmd}, split by
file into training, validation, and test parts (85/5/10) and processed without
text prompts. The activation corpora extracted from these files contain
16,559,916 token states in 32,479 windows of at most 510 tokens for \mmodel{},
with absolute time rebased in each window, and 24,180,079 states in 62,539
windows cut at bar boundaries for \tmodel{}. For characterization, the same
files yield 4,941 excerpts of four quarter notes with at least four notes, up
to four per file, split by file 70/15/15 into discovery, validation, and test
sets. At most 1,000 excerpts per split are used, which gives 2,526 excerpts for
\mmodel{} and 2,498 for \tmodel{}. Discovery excerpts generate hypotheses, validation excerpts filter
them, and test excerpts evaluate them.

\paragraph{Sparse autoencoders.}
We center the states and divide them by the median norm over 50,000 training
tokens, $x=(h-\mu)/s$, because in pilot runs a few tokens with unusually large
norms dominated mean-based scaling and recurred among the top examples of many
features. The SAE is
\begin{equation}
 f(x)=\operatorname{TopK}_{k}\!\left(\operatorname{ReLU}(W_{\rm enc}x+b_{\rm enc})\right),
 \qquad \widehat{x}=W_{\rm dec}f(x)+b_{\rm dec},
 \label{eq:sae}
\end{equation}
where $f_j$ is the activation of feature $j$ and column $j$ of $W_{\rm dec}$
is its direction. With $k=32$, \tmodel{} dictionaries have 6,144 features
(expansion factor 8) and \mmodel{} dictionaries 8,192 (expansion factor 4),
which gives similar sizes at the same sparsity; Appendix~\ref{app:selection}
reports the pilot comparisons behind these choices. We minimize the mean
squared error for 100 epochs with batches of 2,048 tokens and AdamW without
weight decay, using a cosine schedule from $2\times10^{-4}$ to
$2\times10^{-6}$ and gradient clipping at norm 1. Decoder columns are
renormalized after every update, and features that do not fire for ten epochs
are reinitialized toward poorly reconstructed examples
\citep{bricken2023monosemanticity,gao2025scaling}. Each final dictionary is a
single training run with seed 42, and a feature is identified by its model,
layer, and index.

\paragraph{Dictionary quality.}
Following \citet{singh2026discovering}, we retain features that fire in
1--25\% of extraction windows, which leaves between 1,746 and 4,875 features
per dictionary; dead features make up at most 2.42\% of any dictionary
(Table~\ref{tab:health}). Replacing the residual state by its SAE
reconstruction on 200 test excerpts from 146 files recovers 99.22\%
[98.94, 99.51] of the loss for \tmodel{} L13 and 95.57\% [93.95, 97.09] for
\mmodel{} L12, relative to replacing it by its mean,
$\RCE=1-\overline{\Delta\mathrm{CE}}_{\rm SAE}/\overline{\Delta\mathrm{CE}}_{\rm mean}$
\citep{gao2025scaling}, with 95\% intervals from resampling files
(Table~\ref{tab:fidelity}). The other six dictionaries have only the
reconstruction diagnostics.

\begin{table}[t]
\centering\small
\caption{The final dictionaries retain large candidate sets after frequency filtering.
FVU: on the extracted activation corpus; frequent: firing in more than 25\% of
windows; retained: firing in 1--25\%. One training run per row; the models
differ in architecture, tokenization, segmentation, and dictionary size.}
\label{tab:health}
\begin{tabular}{lrrrrrr}
\toprule
Model & Layer & Features & FVU & Dead (\%) & Frequent (\%) & Retained\\
\midrule
text2midi & 2 & 6144 & 0.0311 & 2.18 & 53.22 & 2369 \\
text2midi & 8 & 6144 & 0.0747 & 0.63 & 55.79 & 2143 \\
text2midi & 13 & 6144 & 0.1750 & 0.42 & 63.17 & 1746 \\
text2midi & 16 & 6144 & 0.1568 & 0.16 & 61.12 & 2082 \\
MIDI-LLM & 2 & 8192 & 0.2493 & 2.42 & 22.66 & 3742 \\
MIDI-LLM & 7 & 8192 & 0.3672 & 0.31 & 21.02 & 4422 \\
MIDI-LLM & 12 & 8192 & 0.3865 & 0.23 & 17.50 & 4566 \\
MIDI-LLM & 14 & 8192 & 0.4437 & 0.24 & 17.03 & 4875 \\
\bottomrule
\end{tabular}
\end{table}

%% file: sections/observation.tex
\section{Observational Evidence}
\label{sec:observation}

This section collects the three kinds of observational evidence and applies
their controls. Each kind yields intervention candidates for
Section~\ref{sec:interventions}.

\subsection{Correlations with musical metrics}
\label{sec:associations}

For every characterization excerpt we compute twelve MusPy metrics
\citep{dong2020muspy} of pitch diversity, pitch range, polyphony, scale
consistency, empty beats, and drum-grid alignment (Table~\ref{tab:metrics}),
and we take a feature's activation on an excerpt as its maximum over the
excerpt's tokens. For each feature that passes the window-frequency filter and
is active on 1--50\% of discovery excerpts, we select its most strongly
correlated metric on discovery excerpts and call the association replicated at
threshold $\tau$ if the Pearson correlation reaches $|r|\geq\tau$ with the same
sign on discovery, validation, and test excerpts. Replicated associations are
common: at $\tau=0.20$, 32 to 193 features per dictionary pass discovery and
validation, and 72--81\% of them also pass on test excerpts
(Table~\ref{tab:associations}). Most concern pitch count, polyphony, pitch
range, and pitch entropy (Table~\ref{tab:metric_prevalence}).

These metrics can grow with the number of notes. We therefore regress both the
activation and the metric on $\log(1+\text{notes})$ and $\log(1+\text{tokens})$,
with coefficients fitted on discovery excerpts, and apply the replication
criterion to the residuals. This reanalysis covers all features of 21
dictionaries, the eight final ones and 13 earlier checkpoints, including
features that the window-frequency filter excludes, so its counts are compared
only with each other. The adjustment removes most associations
(Table~\ref{tab:density}). At $\tau=0.20$, the eight final dictionaries contain
10,018 replicated raw associations and 862 adjusted ones, so each dictionary
keeps between 3.4\% and 12.5\% of its raw count (3.4--7.2\% in \mmodel{} and
8.7--12.5\% in \tmodel{}); the earlier checkpoints show the same reduction
(Table~\ref{tab:density_all}). Of the 862 adjusted associations, 376 concern
polyphony, 127 pitch count, 113 the duple drum grid, and 4 pitch entropy. The
strongest adjusted associations supply the density-adjusted candidates of
Section~\ref{sec:protocol}.

\begin{table}[t]
\centering\small
\caption{Adjusting for note and token counts removes most replicated metric
associations. Counts: features whose strongest discovery metric reaches
$|r|\geq\tau$ with the same sign on discovery, validation, and test segments,
before (raw) and after (adj.) residualizing on log counts; eligible: firing on
1--50\% of discovery segments; kept: adj./raw at $\tau=0.20$. Earlier
checkpoints: Appendix~\ref{app:density}.}
\label{tab:density}
\begin{tabular}{lrrrrrrr}
\toprule
& & & \multicolumn{2}{c}{$\tau=0.10$} & \multicolumn{2}{c}{$\tau=0.20$} & \\
\cmidrule(lr){4-5}\cmidrule(lr){6-7}
Model & Layer & Eligible & Raw & Adj. & Raw & Adj. & Kept (\%)\\
\midrule
text2midi & 2 & 5092 & 3295 & 1146 & 1715 & 149 & 8.7 \\
text2midi & 8 & 4724 & 3321 & 1356 & 1773 & 221 & 12.5 \\
text2midi & 13 & 4854 & 3469 & 1358 & 1954 & 199 & 10.2 \\
text2midi & 16 & 5208 & 3565 & 1396 & 1730 & 150 & 8.7 \\
MIDI-LLM & 2 & 4057 & 1773 & 534 & 527 & 38 & 7.2 \\
MIDI-LLM & 7 & 3610 & 1774 & 495 & 839 & 47 & 5.6 \\
MIDI-LLM & 12 & 3117 & 1621 & 329 & 754 & 26 & 3.4 \\
MIDI-LLM & 14 & 3367 & 1689 & 268 & 726 & 32 & 4.4 \\
\bottomrule
\end{tabular}
\end{table}

\subsection{Automated descriptions}
\label{sec:descriptions}

For each final dictionary we build description cards for 50 features with a
stable metric association ($|r|\geq0.15$ in discovery and $\geq0.10$ in
validation, with the same sign) and 50 without one, so that the atlas also
covers patterns outside the metric set. After excluding cards with too few
examples, the atlas contains 786 features, 399 with and 387 without a stable
association. A card shows four top-activating excerpts, four excerpts with
typical positive activation, and four inactive excerpts matched on note count
and metric values, each as a structural and an instrument-colored piano roll
with summary statistics, instrument names, algorithmic key and chord
estimates, and the largest metric contrasts. A large language model, Claude
Sonnet 5, acts as the reviewer of the cards; this LLM reviewer, blind to the
model, layer, feature, activation values, and control status, describes the
shared pattern and then selects the four of eight shuffled held-out excerpts that match it; four are
active and four are matched inactive excerpts. The description and the
selection are made in the same call, so this task evaluates joint description
and recognition rather than an independently frozen description. We report the mean number $H$ of correct
selections among the four picks; random choice gives $\mathbb{E}[H]=2$ and
$P(H=4)=1/70$. In permutation controls the assignment of activations to excerpts is
shuffled; eight such cards are embedded in the atlas run and 30 form a
separate calibration. An automated audit flags cards whose activation peaks
concentrate on one token or one syntactic token category, or whose active and
inactive excerpts differ in note count by more than a quarter.

\begin{figure}[t]
\centering
\includegraphics[width=0.82\linewidth]{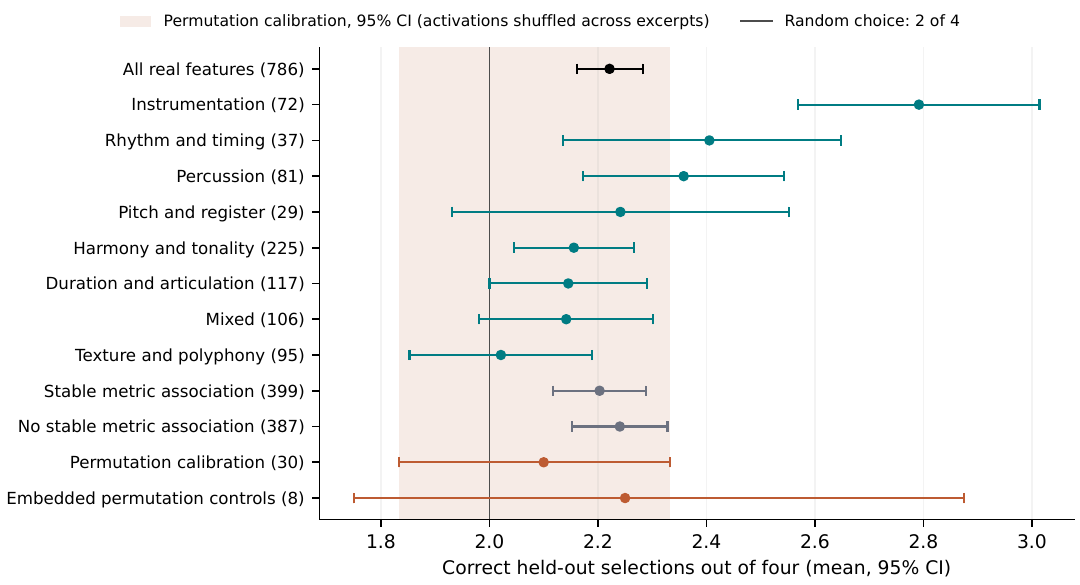}
\caption{Only instrumentation descriptions exceed the permutation calibration.
After describing a feature, the LLM reviewer picks four of eight held-out
excerpts,
four of which activate the feature. Points: mean correct picks, 95\% intervals
resampling cards (group sizes in parentheses); line: random choice (2 of 4);
band: 95\% interval of the 30-card calibration with shuffled activations; form
and unresolved scopes (24 cards) omitted. Values: Table~\ref{tab:recognition}.}
\label{fig:recognition}
\end{figure}

Descriptions recognize held-out excerpts above random choice, with $H=2.221$
$[2.162, 2.282]$, and 48 of 786 cards reach $H=4$, where random choice
predicts 11.2. The permutation controls reach similar values: the LLM reviewer
reports a shared pattern for every control card, the calibration has
$H=2.100$ $[1.833, 2.333]$, and the difference from real cards is $0.121$
$[-0.131, 0.388]$ (Figure~\ref{fig:recognition}). Instrumentation is the
exception. Its 72 cards reach $H=2.792$ $[2.569, 3.014]$, exceed the
calibration by $0.692$ $[0.364, 1.033]$, and provide 21 of the 48 perfect
cards. For every other group, including harmony and tonality (225 cards,
$H=2.156$) and texture and polyphony ($H=2.021$), the interval for the
difference includes zero, and features with and without a stable metric
association have similar values, $2.203$ and $2.240$. The audit flags 504 of the 786 cards, and 12 of the 48 perfect cards
pass it. Table~\ref{tab:description_examples} lists audit-passing
descriptions with perfect recognition, which range from instrument identity to
articulation and harmonic stasis.

\begin{table}[t]
\centering\small
\caption{Selected audit-passing descriptions identify all four active held-out
excerpts. Labels are the LLM reviewer's short labels from the archived
responses.}
\label{tab:description_examples}
\begin{tabularx}{\linewidth}{lXc}
\toprule
Feature & Description & Recognition\\
\midrule
text2midi L13 f3340 & Synth-lead (Lead 2 sawtooth) / electronic-patch instrumentation & 4/4 \\
MIDI-LLM L7 f499 & Solo acoustic-piano melody & 4/4 \\
MIDI-LLM L2 f5880 & Short, detached notes with thin overlap & 4/4 \\
MIDI-LLM L2 f1732 & Dense, drum-heavy texture with chromatic/out-of-scale notes & 4/4 \\
MIDI-LLM L2 f793 & Wide, harmonically static chord hits with weak tonal clarity & 4/4 \\
\bottomrule
\end{tabularx}
\end{table}

\subsection{Shared instrument programs}
\label{sec:programs}

Both vocabularies encode the instrument program of every note, so a feature
whose active excerpts share a program yields a hypothesis that can be measured
directly in the output. Twenty-seven atlas features, 23 in \tmodel{} and 4 in
\mmodel{}, have a General MIDI program that occurs in at least 80\% of their
active discovery excerpts and in three of four held-out active excerpts, and in
none of the matched inactive excerpts. One feature, \tmodel{} L13 f3340, is
not among them, since only its top-activating excerpts contain its program; it
enters the interventions through its description, ``Synth-lead (Lead 2
sawtooth) / electronic-patch instrumentation'', and we call it the synth-lead
feature. Every top-activating discovery excerpt of this feature contains
program 81, \emph{Lead 2 (sawtooth)}, counting programs from
zero, whereas its typical-activation excerpts, at about a tenth of the top
values, and its matched inactive excerpts contain none. On the held-out test, all four active excerpts
contain program 81 and none of the four distractors does.
\FloatBarrier

%% file: sections/interventions.tex
\section{Interventions During Generation}
\label{sec:interventions}

\subsection{Protocol}
\label{sec:protocol}

\paragraph{Candidates.}
A targeted screen tests the three pools of Section~\ref{sec:observation}: for
each model and metric, the feature with the largest minimum of its absolute
discovery and validation correlations (24 feature--metric hypotheses involving
15 features, Table~\ref{tab:muspy_best}); the 27 program features; and the 48
features whose descriptions identify all four held-out active excerpts, of
which the 12 that pass the audit form the primary subset. Ten features belong
to both the program and the description pool, which leaves 80 candidates
(Table~\ref{tab:candidate_pools}). A separate density-adjusted screen takes two
features from each of the 21 dictionaries whose adjusted correlation reaches
$0.10$ in discovery and validation with the same sign: the strongest
association and a second one on another metric, preferably without a close
match in other dictionaries of the same layer. Its metrics are polyphony and
polyphony rate in \mmodel{} and polyphony and pitch range in \tmodel{}.

\begin{table}[t]
\centering\footnotesize\setlength{\tabcolsep}{3pt}
\caption{Three observational routes define the intervention pools. The pools
overlap and contain 80 distinct features.}
\label{tab:candidate_pools}
\begin{tabularx}{\linewidth}{l>{\raggedright\arraybackslash}Xcc>{\raggedright\arraybackslash}X}
\toprule
Pool & Selection & Features & Hypotheses & Evaluation\\
\midrule
MusPy & The strongest discovery/validation association for each metric and model. & 15 & 24 & The metric changes in the direction predicted by its correlation. \\
MIDI programs & The program is frequent in active examples and absent from matched inactive examples. & 27 & 27 & The target program occurs; note and duration shares describe its contribution. \\
Descriptions & The description identifies all four active test excerpts; twelve cards pass the audit. & 48 & 48 & Blinded ratings measure agreement with the fixed description. \\
\bottomrule
\end{tabularx}
\end{table}

\paragraph{Intervention.}
For the unit decoder direction $d_j=W_{{\rm dec},:,j}/\|W_{{\rm dec},:,j}\|_2$
we set
\begin{equation}
 h'_{\ell,t}=h_{\ell,t}+s\,\Delta f_j\,d_j,\qquad
 \Delta f_j=\max\{0,f^\star-f_j((h_{\ell,t}-\mu)/s)\},
 \label{eq:intervention}
\end{equation}
where the target $f^\star$ is the 50th, 75th, or 90th percentile ($P_{50}$,
$P_{75}$, $P_{90}$) of the feature's positive excerpt maxima in discovery. The
edit applies to every consumed MIDI state; prompt processing and the state that
produces the first MIDI token are unchanged. Because the encoder, decoder, and
TopK selection are not mutually inverse, re-encoding $h'$ need not return
exactly $f^\star$. The control replaces $d_j$ by a fixed random unit direction
orthogonal to it and applies the same update rule on its own trajectory.
Ablation sets $\Delta f_j=-f_j$ at naturally active states and has its own
random control.

\paragraph{Design.}
The targeted screen evaluates each feature on ten prompts with paired seeds
under nine conditions, a baseline plus three targets and the ablation with
their controls, which gives 7,200 records; baselines are shared, so the
records are not independent. Generation stops at 512 tokens for \tmodel{} and
768 for \mmodel{}. The prespecified primary contrast is feature minus random
direction at $P_{90}$. A metric hypothesis is tested on the difference
oriented by the sign of the correlation with a paired sign-flip test. A
program hypothesis is tested on the presence of at least one non-drum note
with the program, using an exact one-sided test on prompts where only one
condition contains it; the program's note share is a secondary endpoint.
Benjamini--Hochberg correction \citep{benjamini1995controlling} applies within
each endpoint family at $q<0.05$, and intervals resample prompt pairs with
20,000 draws. The density-adjusted screen uses eight new prompts, a new seed,
512 tokens, 1,000 bootstrap draws, and the baseline, $P_{50}$, $P_{90}$, and ablation conditions with
their controls, with the same primary contrast corrected across its 42
features. Descriptions are tested by blinded ratings of the LLM reviewer
(Section~\ref{sec:semantic}).

\paragraph{Screens and confirmation.}
With eight or ten prompts per feature, the exact per-feature tests have coarse
resolution (Appendix~\ref{app:screen}). We therefore use the screens for two
purposes. Their per-feature effects rank candidates for confirmation on new
prompts, the route that the synth-lead feature took after an earlier
exploratory run (Section~\ref{sec:f3340}). Their family-level effects estimate
whether the features selected by one kind of evidence move their targets: each
oriented effect is divided by the standard deviation of the baseline outputs
over the same prompts, the standardized effects are averaged over the family,
and 95\% intervals resample the prompts, which all features of a screen share,
with 20,000 draws. Appendix~\ref{app:screen} reports sensitivity analyses. We
specified the family-level analyses after the screens.

\paragraph{Manipulation check.}
At $P_{90}$, the edit changes the residual state by $21.4\pm16.2$\% of its norm
in \tmodel{} and $12.9\pm10.8$\% in \mmodel{}, and the random direction by a
similar amount. Re-encoding activates the target feature in $55.8$\% of edited
\tmodel{} states and $35.9$\% of \mmodel{} states, against at most 0.3\% under
the random direction, and all 7,200 records are valid MIDI files with similar
note counts across conditions (Table~\ref{tab:diagnostics}). Features selected
on unprompted Lakh excerpts rarely fire during prompted generation: in the
density-adjusted screen, 18 of the 22 \mmodel{} features never fire at
baseline, and 4 do not fire after re-encoding the $P_{90}$ edit. Removing
their contribution usually leaves the output unchanged, so the experiments
test whether a direction suffices to change the output and leave its necessity
open.

\subsection{Confirmation of an instrument feature}
\label{sec:f3340}

The synth-lead feature (f3340) was among the features of an exploratory
intervention run carried out
before the description stage, and its description was written without access
to that run. After the description was revealed, the earlier generations
served as a pilot: program 81 occurred in 4 of 10 feature-direction outputs at
$P_{90}$ and in no baseline or random output. We then prespecified the
hypothesis that inducing the feature at $P_{90}$ increases the presence of program
81 relative to the random direction, together with the endpoint, the
contrast, 20 new prompts, the seed, and four success criteria: a significant
paired test, prevalence above baseline, a positive effect in both halves of
the prompt list, and a manipulation check on the re-encoded activation.

\begin{table}[t]
\centering\small\setlength{\tabcolsep}{4pt}
\caption{The program-81 effect of the synth-lead feature (\tmodel{} L13 f3340) recurs on two disjoint prompt
sets at $P_{90}$. Counts: outputs with at least one non-drum program-81 note;
lift: feature minus random presence in percentage points, paired-prompt
bootstrap 95\% interval; $p$: exact one-sided test on discordant pairs. The
confirmation was prespecified.}
\label{tab:f3340}
\begin{tabular}{lrrrrrr}
\toprule
Study & Prompts & Baseline & Feature & Random & Lift (pp) [95\% CI] & $p$\\
\midrule
Confirmation & 20 & 2 & 14 & 1 & 65 [45, 85] & 0.000122 \\
Replication & 10 & 0 & 6 & 0 & 60 [30, 90] & 0.015625 \\
\bottomrule
\end{tabular}
\end{table}

The confirmation passes all four criteria (Table~\ref{tab:f3340}). Program 81
occurs in 14 of 20 feature-direction outputs, 1 of 20 random-direction
outputs, and 2 of 20 baselines; thirteen prompt pairs contain it only under
the feature direction and none only under the random direction
($p=0.000122$). The targeted screen replicates the effect on ten disjoint
prompts with a different seed: program 81 occurs in 6 of 10 feature-direction
outputs and in no baseline or random output ($p=0.0156$). Presence grows with
the target, from 0 of 10 at $P_{50}$ to 4 of 10 at $P_{75}$ and 6 of 10 at
$P_{90}$, where the program's share of non-drum notes rises by 6.6 percentage
points $[2.9, 10.4]$ (Figure~\ref{fig:dose}a); the atlas contains every output
of this replication. The feature also reappears in independently trained
dictionaries. Matching each feature to the most correlated feature of another
dictionary at the same layer, on discovery segments with evaluation on
validation segments, the synth-lead feature has a counterpart in an earlier \tmodel{} L13
dictionary with the same expansion factor ($r=0.970$ and $0.943$) and in one
with expansion factor 4 ($r=0.980$ and $0.962$), whereas only 35.2\% and
25.6\% of the final L13 features have a match with $r\geq0.8$ on both splits.

The endpoint is the program number written in the MIDI file, so the result
shows that the feature direction controls which program the model selects.

\subsection{Family-level effects}
\label{sec:screen}

Table~\ref{tab:outcomes} summarizes all hypothesis families.

\begin{table}[t]
\centering\footnotesize\setlength{\tabcolsep}{3.5pt}
\caption{Intervention outcomes by hypothesis source. Test: $p$ of a
prespecified test; per-feature screen results are in Appendix~\ref{app:screen}.
Family level: mean oriented effect at $P_{90}$ in baseline SD, 95\%
interval over shared prompts (Table~\ref{tab:family}); program row: features
whose program is absent from all baselines; description row: pooled LLM rating
test of the primary subset.}
\label{tab:outcomes}
\begin{tabularx}{\linewidth}{>{\raggedright\arraybackslash}p{0.19\linewidth}>{\raggedright\arraybackslash}p{0.1\linewidth}rrl>{\raggedright\arraybackslash}X}
\toprule
Hypothesis source & Endpoint & Hyp. & Prompts & Test & Family-level result\\
\midrule
Synth-lead feature, confirmation & Program presence & 1 & 20 & $p=0.000122$ & Prespecified; passes all four criteria \\
Synth-lead feature, replication & Program presence & 1 & 10 & $p=0.0156$ & Disjoint prompts and seed \\
Density-adjusted correlation, 21 SAEs & MusPy metric & 42 & 8 & -- & $+0.18$ SD $[-0.04, +0.37]$ vs.\ baseline; $+0.17$ SD $[+0.02, +0.33]$ vs.\ random \\
Program frequency & Program presence & 27 & 10 & -- & Absent programs introduced for 6 of 10 features (random: 1) \\
Strongest raw metric correlation & MusPy metric & 24 & 10 & -- & $-0.23$ SD $[-1.23, +0.91]$ vs.\ random \\
Description, audit-passing & Blinded LLM rating & 12 & 5 & $p=0.3125$ & Synth-lead feature rated highest; pooled interval includes 0 \\
\bottomrule
\end{tabularx}
\end{table}

\begin{table}[t]
\centering\footnotesize\setlength{\tabcolsep}{3pt}
\caption{Family-level effects. Mean oriented effect in baseline SD with 95\%
interval resampling the shared prompts; positive values agree with the
correlation. Positive: hypotheses with feature $-$ random $>0$; no $P_{75}$
condition in the density-adjusted screen.}
\label{tab:family}
\begin{tabular}{llrrrr}
\toprule
Family (hypotheses, SAEs) & Target & Feature $-$ random & Positive & Feature $-$ baseline & Random $-$ baseline\\
\midrule
Density-adjusted (42, 21 SAEs) & $P_{50}$ & $+0.02$ $[-0.09, +0.12]$ & 22/42 & $-0.01$ $[-0.16, +0.12]$ & $-0.04$ $[-0.16, +0.10]$ \\
 & $P_{90}$ & $+0.17$ $[+0.02, +0.33]$ & 27/42 & $+0.18$ $[-0.04, +0.37]$ & $+0.01$ $[-0.15, +0.19]$ \\
Raw correlation (24, 7 SAEs) & $P_{50}$ & $-0.04$ $[-0.30, +0.31]$ & 9/24 & $+0.14$ $[-0.18, +0.53]$ & $+0.18$ $[-0.11, +0.48]$ \\
 & $P_{75}$ & $+0.15$ $[-0.55, +1.01]$ & 15/24 & $+0.06$ $[-0.26, +0.44]$ & $-0.09$ $[-0.70, +0.56]$ \\
 & $P_{90}$ & $-0.23$ $[-1.23, +0.91]$ & 8/24 & $-0.24$ $[-1.20, +0.51]$ & $-0.02$ $[-0.72, +0.54]$ \\
\bottomrule
\end{tabular}
\end{table}

\paragraph{Density-adjusted metric features.}
At $P_{90}$, the 42 density-adjusted features move their target metric in the
predicted direction by $0.17$ baseline standard deviations $[0.02, 0.33]$
relative to the random direction and by $0.18$ $[-0.04, 0.37]$ relative to
baseline generation, while the random direction itself barely changes it
($+0.01$ $[-0.15, 0.19]$; Table~\ref{tab:family}). The shift appears at
$P_{90}$ and not at $P_{50}$, and the estimates relative to baseline are
similar in the two models, $0.20$ $[0.07, 0.32]$ for \mmodel{} and $0.15$
$[-0.23, 0.49]$ for \tmodel{}. Of the 42 oriented effects, 27 are positive.
The largest, a polyphony feature from a pilot dictionary (\tmodel{} L2
f5056), raises polyphony by 1.63 $[0.68, 2.58]$ relative to the random
direction (Table~\ref{tab:robustness}).

\begin{figure}[t]
\centering
\includegraphics[width=\linewidth]{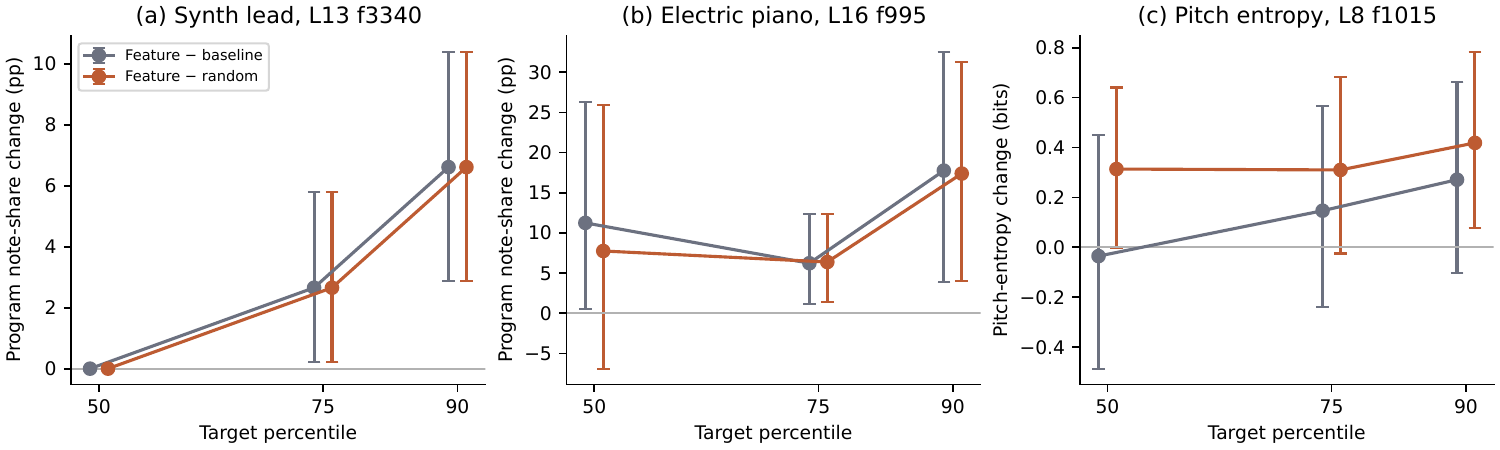}
\caption{Dose responses of selected \tmodel{} features relative to the baseline and
the matched random direction. Points: paired mean differences over ten
prompts, 95\% bootstrap intervals; (a, b) target-program share of non-drum
notes, (c) pitch entropy. In (a), the replication for the synth-lead feature, baseline and random
outputs contain no program-81 notes, so the two comparisons coincide; (b) and
(c) are the largest exploratory effects of their families. $P_{90}$ is the
prespecified dose.}
\label{fig:dose}
\end{figure}

\paragraph{Raw-correlation metric features.}
Features selected by raw correlations show no such shift. At $P_{90}$ the
estimate is $-0.23$ $[-1.23, 0.91]$ relative to the random direction, and 8 of
the 24 effects have the predicted sign (Table~\ref{tab:all_muspy}). The
largest single effect belongs to a pitch-entropy feature (\tmodel{} L8 f1015),
which exceeds the random direction by 0.418 bits of pitch entropy
$[0.078, 0.783]$ and the baseline by 0.271 bits $[-0.102, 0.661]$, and its
difference from the baseline grows across the three targets
(Figure~\ref{fig:dose}c).

\paragraph{Program features.}
The program features show no consistent increase in presence: the feature direction
yields more program-positive outputs than the random direction for 10 of the
27 features and fewer for 10 (Table~\ref{tab:all_instruments}). In an
exploratory comparison restricted to the ten features whose program occurs in
no baseline output, the situation of the synth-lead feature, the feature
direction introduces
the program for six features and the random direction for one; eight prompts
contain the program only under the feature direction and one only under the
random direction. The largest note-share increase belongs to an
electric-piano feature (\tmodel{} L16 f995, program 4, \emph{Electric Piano
1}). The program occurs in 5 of 10
feature-direction outputs and in 1 of 10 baseline and random outputs,
and its note share rises by 17.4 percentage points over the random direction
$[4.0, 31.2]$, concentrated in four
prompts (Figure~\ref{fig:paired_prompts}). The electric-piano and
pitch-entropy features are the strongest candidates for confirmation.

\subsection{Baseline and random references}
\label{sec:two_references}

Every effect is read against both references, because the random direction
can itself degrade the target property. For a bright-piano feature (\tmodel{} L2 f571), program 1, \emph{Bright
Acoustic Piano}, occurs in 7 of 10 baselines,
and both directions reduce its note share, the feature direction less so. For
a polyphony-rate feature (L13 f3827) the polyphony rate exceeds the random direction by 0.132 but differs
from the baseline by $-0.003$, and for a pitch-class feature (L8 f5166) the pitch-class count exceeds
the random direction by 1.10 classes and the baseline by only 0.10
(Table~\ref{tab:muspy_selected}); compared with the random direction alone,
these would appear as successes. The interventions also change untargeted
properties: for 12 of the 15 metric features, the largest standardized change
occurs on a metric other than the hypothesized one
(Figure~\ref{fig:cross_metric}).

\subsection{Blinded LLM ratings of descriptions}
\label{sec:semantic}

Metric and program endpoints cannot test descriptions of texture,
articulation, or harmony. For each of the 48 description candidates, the same
LLM reviewer therefore rates five generated prompt triplets. Each triplet shows
the baseline as a reference and the feature and random outputs as anonymous
variants, as structural and instrument-colored piano rolls; the LLM reviewer
rates
how well each variant matches the fixed description on a scale from 1 to 5
and may pick the better-matching one. Differences are averaged within features
and tested across features with a bootstrap and a sign-flip test
(Appendix~\ref{app:semantic_features}, Table~\ref{tab:semantic}). In the audit-passing subset, the feature direction
is rated 0.167 points higher than the random direction $[-0.300, 0.700]$
($p=0.3125$), and across all 48 descriptions 0.108 points higher
$[-0.112, 0.338]$, with the feature variant identified in 16 of 34
directional choices. The synth-lead feature has the largest difference, $+2.40\pm0.55$, and is
rated higher on all five prompts; without it, the primary subset has a
difference of $-0.036$ $[-0.400, 0.291]$. The next five largest differences
concern percussion, multi-instrument texture, and harmonic rhythm, and each of
these features carries an audit flag (Appendix~\ref{app:semantic_features}).

\FloatBarrier

%% file: sections/discussion.tex
\section{Discussion}
\label{sec:discussion}

\paragraph{The role of the controls.}
The density adjustment and the permutation calibration separate observational
evidence that points to steerable features from evidence that does not.
Features selected after the density adjustment show a small positive
family-level shift of their target metric, whereas features selected by raw
correlations show none; because the two screens also differ in prompts,
dictionaries, and metrics, this contrast describes the two selections rather
than isolating the effect of the adjustment. The instrumentation descriptions,
the one group that exceeds the permutation calibration, yield a feature that
controls its instrument in a prespecified confirmation and a replication. We
therefore recommend applying both controls before selecting features.

\paragraph{Instrument programs and explicit tokens.}
The synth-lead feature concerns an attribute that both vocabularies encode
explicitly: REMI+ writes the instrument as a separate event, and AMT encodes
instrument and pitch jointly in each note token. A direction tied to program
81 can change the output by raising the probability of a small set of tokens,
whereas polyphony, pitch range, and pitch entropy are statistics over many
token decisions. This suggests a selection criterion that the protocol can
add: features whose decoder directions project strongly onto a few output
tokens should steer more often than features selected by input correlations,
as \citet{arad2025saes} and \citet{gurarieh2025enhancing} report for language
models; output projections that did not separate the synth-lead feature from
the other candidates would refute this hypothesis. The program route also
points to further candidates: in an exploratory comparison, program features
introduced programs absent from baseline generation more often than random
directions did.

\paragraph{Unsupervised and supervised directions.}
Our earlier study steered the same models with supervised difference-in-means
directions and evaluated them in both orientations, which separates
directional responses from symmetric drift \citep{pocwiardowski2026mimidi};
the present study likewise needs a second reference. The present study complements
it in two ways. The SAE exposed a controllable feature for an attribute that
was not chosen in advance, in line with \citet{peng2026position}. The
description route also found a feature that the program-frequency criterion of
Section~\ref{sec:programs} did not select, because the program occurs in its
strongest activations and not in its typical ones; combining several routes
therefore widens the set of candidates.

\paragraph{Recommendations.}
For SAE steering studies of symbolic generators, the protocol suggests five
practices: adjust metric correlations for note and token counts before
interpreting them or selecting candidates; compare automated descriptions with
permutation controls run in the same pipeline; compare interventions with both
a baseline and a random direction; use screens with few prompts per feature to
rank candidates and estimate family-level effects, and confirm individual
features on new prompts with a fixed endpoint; and check that a candidate
reappears in independently trained dictionaries.

\paragraph{Limitations.}
Each final dictionary is a single training run, and the cross-dictionary match
supports the stability of the synth-lead feature but not of whole
dictionaries. The screens use
eight or ten prompts and one random direction per candidate. The family-level
estimates were specified after the screens, average standardized effects
across metrics, and depend on the resampling scheme
(Appendix~\ref{app:screen}). The LLM reviewer
is Claude Sonnet 5 without recorded sampling settings; it writes each
description and makes the held-out choice in the same call, sees instrument
names and summary statistics, which may contribute to the recognition of
instrumentation, and also rates the generated music. All
endpoints are symbolic or visual, and we did not run a listening study. The
SAE training and characterization splits were drawn independently from the
same files without a file-level overlap audit, so held-out refers to the
characterization split. The two models differ in architecture, tokenization,
pretraining, and dictionary configuration, so we do not attribute differences
between them to any one factor. The endpoint of the synth-lead confirmation is the program
number in MIDI, so the result does not establish a representation of perceived
timbre. Interventions use one site per layer and targets derived from
unprompted excerpts.

\section{Conclusion}

We asked whether observational evidence about SAE features predicts control in
text-to-MIDI generation. A feature identified through its description, in the
one description group that survives the permutation control, steered its
instrument program in a prespecified confirmation and a replication and
reappeared in independently trained dictionaries. Features selected by
density-adjusted correlations showed a small positive family-level shift of
their target metric, and features selected by raw correlations showed none.
The controls matter because most raw correlations reflect note density and
most descriptions match permutation controls. Screens rank candidates and
estimate family-level effects, and confirmation on new prompts establishes
individual features. The atlas at
\url{https://jpocwiar.github.io/MI-MIDIv2-Demo/} provides piano rolls, MIDI
files, and synthesized audio for all 80 features of the targeted screen, with
every prompt for ten highlighted features, including the synth-lead feature,
and one prompt selected by its observed effect for each of the others.
Confirming the ranked candidates, measuring output projections of decoder
directions, listening tests, and repeated dictionary seeds are the next steps.

%% file: sections/appendix.tex
\section{Dictionary Selection and Fidelity}
\label{app:selection}

Table~\ref{tab:pilots} reports the pilot runs behind the choice of expansion
factor (EF) and TopK budget $k$. In \mmodel{} L12, EF 32 at $k=32$ left
83.72\% of the dictionary dead, and $k=128$ improved reconstruction and reduced
dead features at the cost of more active features per token; we kept EF 4 and
$k=32$ as a compromise between reconstruction, sparsity, and cost. In
\tmodel{}, EF 8 retained more features after frequency filtering and had
slightly lower FVU than EF 4 at both pilot layers. The nine pilot runs and four
earlier runs of the final configurations (\mmodel{} L2 and L7, \tmodel{} L8
and L16) form the 13 earlier checkpoints of the density reanalysis and the
density-adjusted screen. Table~\ref{tab:fidelity} reports the functional
fidelity of the two evaluated final dictionaries.

\begin{table}[htbp]
\centering\small
\caption{Pilot comparisons for the expansion factor (EF) and TopK budget $k$. Dead and
frequent: window-frequency definitions of Table~\ref{tab:health}; one pilot run
per row.}
\label{tab:pilots}
\begin{tabular}{lrrrrrrr}
\toprule
Model & Layer & EF & $k$ & FVU & Dead (\%) & Frequent (\%) & Retained\\
\midrule
text2midi & 2 & 4 & 32 & 0.0370 & 0.65 & 80.57 & 496 \\
text2midi & 2 & 8 & 32 & 0.0316 & 1.56 & 55.76 & 2174 \\
text2midi & 13 & 4 & 32 & 0.2038 & 0.13 & 83.98 & 288 \\
text2midi & 13 & 8 & 32 & 0.1885 & 0.13 & 67.04 & 1489 \\
MIDI-LLM & 12 & 4 & 32 & 0.3693 & 0.43 & 18.88 & 4968 \\
MIDI-LLM & 12 & 32 & 32 & 0.3923 & 83.72 & 1.84 & 2822 \\
MIDI-LLM & 12 & 32 & 128 & 0.2266 & 30.29 & 3.05 & 6566 \\
MIDI-LLM & 14 & 4 & 32 & 0.4242 & 0.11 & 19.60 & 5247 \\
MIDI-LLM & 14 & 32 & 128 & 0.2731 & 34.46 & 3.59 & 5745 \\
\bottomrule
\end{tabular}
\end{table}

\begin{table}[htbp]
\centering\footnotesize\setlength{\tabcolsep}{4pt}
\caption{Functional fidelity of the two evaluated dictionaries. $\Delta\mathrm{CE}$:
increase in next-token cross-entropy over the unmodified model; recovery:
$\RCE$ from file-balanced means on 200 excerpts from 146 files; 95\%
file-bootstrap intervals.}
\label{tab:fidelity}
\begin{tabular}{lrrr}
\toprule
Model and layer & $\Delta\mathrm{CE}_{\rm SAE}$ & $\Delta\mathrm{CE}_{\rm mean}$ & Recovery (\%)\\
\midrule
text2midi L13 & 0.0324 [0.0203, 0.0443] & 4.1488 [4.0473, 4.2532] & 99.22 [98.94, 99.51] \\
MIDI-LLM L12 & 0.0852 [0.0569, 0.1159] & 1.9234 [1.7918, 2.0575] & 95.57 [93.95, 97.09] \\
\bottomrule
\end{tabular}
\end{table}

\section{Metric Characterization}
\label{app:metrics}

Table~\ref{tab:metrics} defines the twelve MusPy measurements. Pitch range is
undefined for drum-only excerpts and drum metrics for excerpts without drums;
such values are left missing, and correlations use pairwise complete data.
Tables~\ref{tab:associations}, \ref{tab:metric_prevalence}, and
\ref{tab:muspy_best} give the replication counts, the counts per metric, and
the strongest feature for each metric and model.

\begin{table}[htbp]
\centering\small
\caption{The twelve MusPy measurements. Scale consistency measures pitch-class
membership in a diatonic scale and does not assess a tonal center.}
\label{tab:metrics}
\begin{tabularx}{\linewidth}{lX}
\toprule
Metric & Definition\\
\midrule
Pitch entropy & Shannon entropy of the MIDI-pitch distribution.\\
Pitch-class entropy & Shannon entropy after reducing pitch modulo twelve.\\
Pitch count & Number of distinct MIDI pitches.\\
Pitch-class count & Number of distinct pitch classes.\\
Pitch range & Difference between the highest and lowest pitched notes.\\
Polyphony & Average number of simultaneous pitches over nonempty time steps.\\
Polyphony rate & Fraction of time steps with more than two active pitches (the library default).\\
Scale consistency & Largest fraction of notes belonging to a tested diatonic scale.\\
Empty-beat rate & Fraction of quarter-note units without notes.\\
Drum grid (duple) & Fraction of drum onsets aligned with the duple grid.\\
Drum grid (triple) & Fraction of drum onsets aligned with the triple grid.\\
Drum consistency & Maximum of the duple-grid and triple-grid drum rates.\\
\bottomrule
\end{tabularx}
\end{table}

\begin{table}[htbp]
\centering\scriptsize\setlength{\tabcolsep}{3pt}
\caption{Replication of metric associations across characterization splits. D,V:
features with the same sign and $|r|\geq\tau$ in discovery and validation;
next column: of these, also on test; percentages: of eligible features
(window-frequency filter, firing on 1--50\% of discovery excerpts).}
\label{tab:associations}
\resizebox{\linewidth}{!}{%
\begin{tabular}{lrrllllll}
\toprule
& & & \multicolumn{2}{c}{$|r|\geq0.10$} & \multicolumn{2}{c}{$|r|\geq0.20$} & \multicolumn{2}{c}{$|r|\geq0.30$}\\
\cmidrule(lr){4-5}\cmidrule(lr){6-7}\cmidrule(lr){8-9}
Model & Layer & Eligible & D,V selected & Test & D,V selected & Test & D,V selected & Test\\
\midrule
text2midi & 2 & 2252 & 963/2252 (42.8\%) & 755/963 & 137/2252 (6.1\%) & 111/137 & 8/2252 (0.4\%) & 6/8 \\
text2midi & 8 & 1963 & 992/1963 (50.5\%) & 820/992 & 193/1963 (9.8\%) & 150/193 & 13/1963 (0.7\%) & 8/13 \\
text2midi & 13 & 1652 & 745/1652 (45.1\%) & 608/745 & 126/1652 (7.6\%) & 97/126 & 7/1652 (0.4\%) & 6/7 \\
text2midi & 16 & 2008 & 949/2008 (47.3\%) & 765/949 & 140/2008 (7.0\%) & 107/140 & 9/2008 (0.4\%) & 4/9 \\
MIDI-LLM & 2 & 2393 & 734/2393 (30.7\%) & 543/734 & 32/2393 (1.3\%) & 23/32 & 3/2393 (0.1\%) & 3/3 \\
MIDI-LLM & 7 & 2064 & 685/2064 (33.2\%) & 529/685 & 88/2064 (4.3\%) & 69/88 & 10/2064 (0.5\%) & 7/10 \\
MIDI-LLM & 12 & 1844 & 640/1844 (34.7\%) & 497/640 & 97/1844 (5.3\%) & 72/97 & 13/1844 (0.7\%) & 9/13 \\
MIDI-LLM & 14 & 2108 & 753/2108 (35.7\%) & 579/753 & 99/2108 (4.7\%) & 72/99 & 13/2108 (0.6\%) & 11/13 \\
\bottomrule
\end{tabular}}
\end{table}

\begin{table}[htbp]
\centering\small
\caption{Associations per metric, summed over the four final dictionaries of each
model (same sign and $|r|\geq\tau$ in discovery and validation; a feature can
count under several metrics). Eligible features per metric: 7,876 (\tmodel{})
and 8,423 (\mmodel{}); drum metrics: 7,871 and 8,366.}
\label{tab:metric_prevalence}
\begin{tabular}{lrrrrrr}
\toprule
& \multicolumn{3}{c}{\tmodel{}} & \multicolumn{3}{c}{\mmodel{}}\\
\cmidrule(lr){2-4}\cmidrule(lr){5-7}
Metric & $0.10$ & $0.20$ & $0.30$ & $0.10$ & $0.20$ & $0.30$\\
\midrule
Pitch entropy & 2093 & 102 & 0 & 1297 & 78 & 8 \\
Pitch-class entropy & 834 & 10 & 0 & 443 & 10 & 1 \\
Pitch count & 3100 & 439 & 29 & 2454 & 272 & 30 \\
Pitch-class count & 1731 & 79 & 0 & 1216 & 57 & 2 \\
Pitch range & 1969 & 115 & 7 & 1232 & 79 & 11 \\
Polyphony & 1892 & 225 & 8 & 1375 & 141 & 24 \\
Polyphony rate & 1098 & 21 & 0 & 748 & 51 & 8 \\
Scale consistency & 46 & 1 & 0 & 9 & 0 & 0 \\
Empty-beat rate & 14 & 4 & 0 & 4 & 0 & 0 \\
Drum grid (duple) & 338 & 35 & 0 & 164 & 8 & 0 \\
Drum grid (triple) & 450 & 10 & 0 & 257 & 4 & 0 \\
Drum consistency & 271 & 15 & 0 & 167 & 9 & 0 \\
\bottomrule
\end{tabular}
\end{table}

\begin{table}[htbp]
\centering\small\setlength{\tabcolsep}{4pt}
\caption{Strongest feature per metric and model, which defines the metric pool (24
hypotheses, 15 features). $r_{\min}$: smaller absolute discovery or validation
correlation, same sign; $r_T$: test correlation, reported after selection.}
\label{tab:muspy_best}
\begin{tabular}{l lrr lrr}
\toprule
& \multicolumn{3}{c}{\tmodel{}} & \multicolumn{3}{c}{\mmodel{}}\\
\cmidrule(lr){2-4}\cmidrule(lr){5-7}
Metric & Feature & $r_{\min}$ & $r_T$ & Feature & $r_{\min}$ & $r_T$\\
\midrule
Pitch entropy & L8 f1015 & 0.279 & +0.258 & L14 f893 & 0.418 & +0.387 \\
Pitch-class entropy & L8 f5166 & 0.231 & +0.211 & L14 f893 & 0.321 & +0.302 \\
Pitch count & L8 f2187 & 0.372 & +0.295 & L14 f893 & 0.486 & +0.435 \\
Pitch-class count & L8 f5166 & 0.286 & +0.259 & L14 f893 & 0.389 & +0.363 \\
Pitch range & L2 f4317 & 0.402 & +0.387 & L7 f6043 & 0.488 & +0.437 \\
Polyphony & L16 f2307 & 0.410 & +0.424 & L7 f6043 & 0.578 & +0.613 \\
Polyphony rate & L13 f3827 & 0.293 & +0.345 & L14 f893 & 0.463 & +0.417 \\
Scale consistency & L16 f1472 & 0.216 & -0.175 & L7 f6633 & 0.125 & -0.063 \\
Empty-beat rate & L2 f947 & 0.262 & +0.361 & L14 f893 & 0.120 & -0.152 \\
Drum grid (duple) & L13 f1002 & 0.286 & -0.381 & L2 f5050 & 0.230 & -0.156 \\
Drum grid (triple) & L13 f2487 & 0.247 & -0.279 & L2 f6981 & 0.265 & -0.146 \\
Drum consistency & L13 f1002 & 0.285 & -0.337 & L2 f5050 & 0.257 & -0.162 \\
\bottomrule
\end{tabular}
\end{table}

\section{Density Reanalysis of the Earlier Checkpoints}
\label{app:density}

The reanalysis evaluates all features of each dictionary that fire on 1--50\%
of discovery segments, using a shared manifest of 4,941 four-beat segments
from the 1,500 files (3,415 discovery, 776 validation, and 750 test segments,
of which at most 1,000 per split are used); these are the characterization
segments of Section~\ref{sec:setup}. Activations and metrics are
residualized on an intercept, $\log(1+\text{notes})$, and
$\log(1+\text{tokens})$, with coefficients fitted on discovery segments.
Table~\ref{tab:density_all} reports the 13 earlier checkpoints.

\begin{table}[htbp]
\centering\footnotesize\setlength{\tabcolsep}{4pt}
\caption{Density reanalysis of the 13 earlier checkpoints. Counts as in
Table~\ref{tab:density}; historical: earlier runs with the stated layer,
expansion factor, and $k$.}
\label{tab:density_all}
\begin{tabular}{llrrrrrrr}
\toprule
& & & \multicolumn{2}{c}{$\tau=0.10$} & \multicolumn{2}{c}{$\tau=0.20$} & \multicolumn{2}{c}{$\tau=0.30$}\\
\cmidrule(lr){4-5}\cmidrule(lr){6-7}\cmidrule(lr){8-9}
Model & Dictionary & Eligible & Raw & Adj. & Raw & Adj. & Raw & Adj.\\
\midrule
text2midi & L2 EF4 k32 (historical) & 2426 & 1790 & 680 & 1144 & 118 & 403 & 18 \\
text2midi & L2 EF8 k32 (historical) & 5041 & 3316 & 1093 & 1713 & 151 & 480 & 24 \\
text2midi & L8 EF8 k32 (historical) & 4510 & 3283 & 1343 & 1840 & 234 & 692 & 40 \\
text2midi & L13 EF4 k32 (historical) & 2032 & 1727 & 743 & 1175 & 135 & 506 & 28 \\
text2midi & L13 EF8 k32 (historical) & 4763 & 3571 & 1365 & 2024 & 217 & 684 & 41 \\
text2midi & L16 EF8 k32 (historical) & 5029 & 3529 & 1314 & 1629 & 161 & 446 & 26 \\
MIDI-LLM & L2 EF4 k32 (historical) & 4138 & 1787 & 537 & 548 & 52 & 110 & 9 \\
MIDI-LLM & L7 EF4 k32 (historical) & 3777 & 1897 & 527 & 850 & 63 & 332 & 10 \\
MIDI-LLM & L12 EF32 k128 (historical) & 3346 & 1571 & 299 & 792 & 39 & 374 & 4 \\
MIDI-LLM & L12 EF32 k32 (historical) & 2360 & 1251 & 273 & 685 & 31 & 237 & 2 \\
MIDI-LLM & L12 EF4 k32 (historical) & 3451 & 1811 & 356 & 803 & 33 & 243 & 4 \\
MIDI-LLM & L14 EF32 k128 (historical) & 3394 & 1665 & 254 & 818 & 20 & 294 & 4 \\
MIDI-LLM & L14 EF4 k32 (historical) & 3695 & 1938 & 313 & 789 & 28 & 226 & 3 \\
\bottomrule
\end{tabular}
\end{table}

\section{Description Recognition by Group}
\label{app:recognition}

Table~\ref{tab:recognition} gives the values behind
Figure~\ref{fig:recognition}. Expected perfect cards under random choice are
the number of cards divided by 70. Intervals resample cards with 20,000 draws,
the difference resampling both sets independently, and treat cards as
independent, although cards share the LLM reviewer and some source excerpts.

\begin{table}[htbp]
\centering\footnotesize\setlength{\tabcolsep}{4pt}
\caption{Description recognition by group. $H$: mean correct selections out of
four (random choice: 2); difference: to the 30-card permutation calibration;
$H=4$: perfect cards, with the random-choice expectation in parentheses.}
\label{tab:recognition}
\begin{tabular}{lrllr}
\toprule
Group & Cards & $H$ [95\% CI] & Difference to calibration [95\% CI] & $H=4$\\
\midrule
All real features & 786 & $2.221$ $[2.162, 2.282]$ & $+0.121$ $[-0.131, +0.388]$ & 48 (11.2) \\
Instrumentation & 72 & $2.792$ $[2.569, 3.014]$ & $+0.692$ $[+0.364, +1.033]$ & 21 (1.0) \\
Rhythm and timing & 37 & $2.405$ $[2.135, 2.649]$ & $+0.305$ $[-0.057, +0.659]$ & 1 (0.5) \\
Percussion & 81 & $2.358$ $[2.173, 2.543]$ & $+0.258$ $[-0.048, +0.577]$ & 5 (1.2) \\
Pitch and register & 29 & $2.241$ $[1.931, 2.552]$ & $+0.141$ $[-0.239, +0.549]$ & 2 (0.4) \\
Harmony and tonality & 225 & $2.156$ $[2.044, 2.267]$ & $+0.056$ $[-0.211, +0.338]$ & 11 (3.2) \\
Duration and articulation & 117 & $2.145$ $[2.000, 2.291]$ & $+0.045$ $[-0.240, +0.338]$ & 3 (1.7) \\
Mixed & 106 & $2.142$ $[1.981, 2.302]$ & $+0.042$ $[-0.253, +0.345]$ & 2 (1.5) \\
Texture and polyphony & 95 & $2.021$ $[1.853, 2.189]$ & $-0.079$ $[-0.372, +0.230]$ & 3 (1.4) \\
Stable metric association & 399 & $2.203$ $[2.118, 2.288]$ & $+0.103$ $[-0.156, +0.378]$ & 22 (5.7) \\
No stable metric association & 387 & $2.240$ $[2.152, 2.328]$ & $+0.140$ $[-0.119, +0.415]$ & 26 (5.5) \\
Permutation calibration & 30 & $2.100$ $[1.833, 2.333]$ &  & 0 (0.4) \\
Embedded permutation controls & 8 & $2.250$ $[1.750, 2.875]$ &  & 1 (0.1) \\
\bottomrule
\end{tabular}
\end{table}
\FloatBarrier

\section{Complete Intervention Screens}
\label{app:screen}

\begin{figure}[htbp]
\centering
\includegraphics[width=\linewidth]{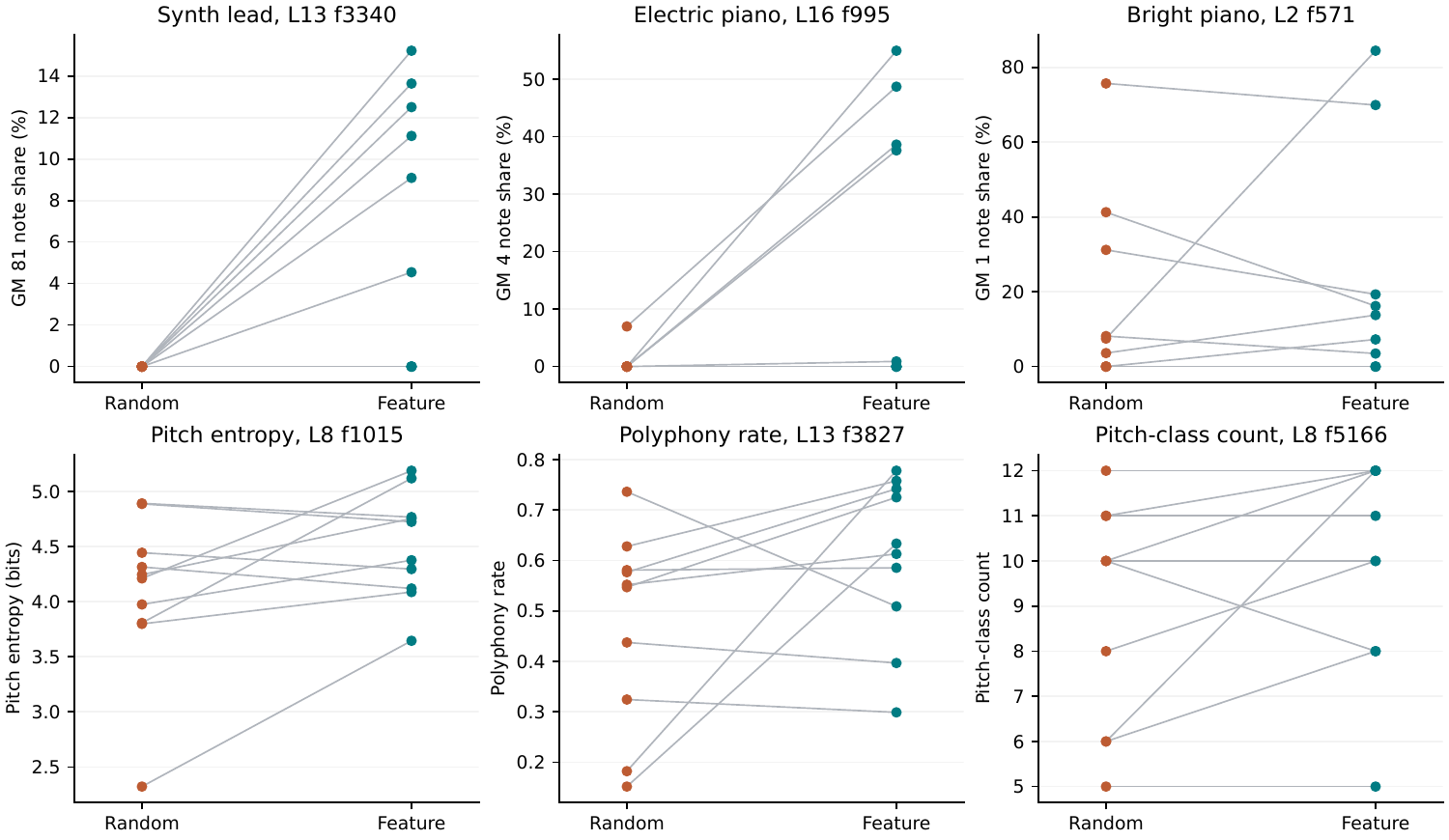}
\caption{Paired prompts at $P_{90}$ for selected effects. Lines join the random and
feature outputs of one prompt; top: target-program note share; bottom:
hypothesized metric; synth lead: replication of the confirmed effect, the others
exploratory.}
\label{fig:paired_prompts}
\end{figure}

\begin{figure}[htbp]
\centering
\includegraphics[width=\linewidth]{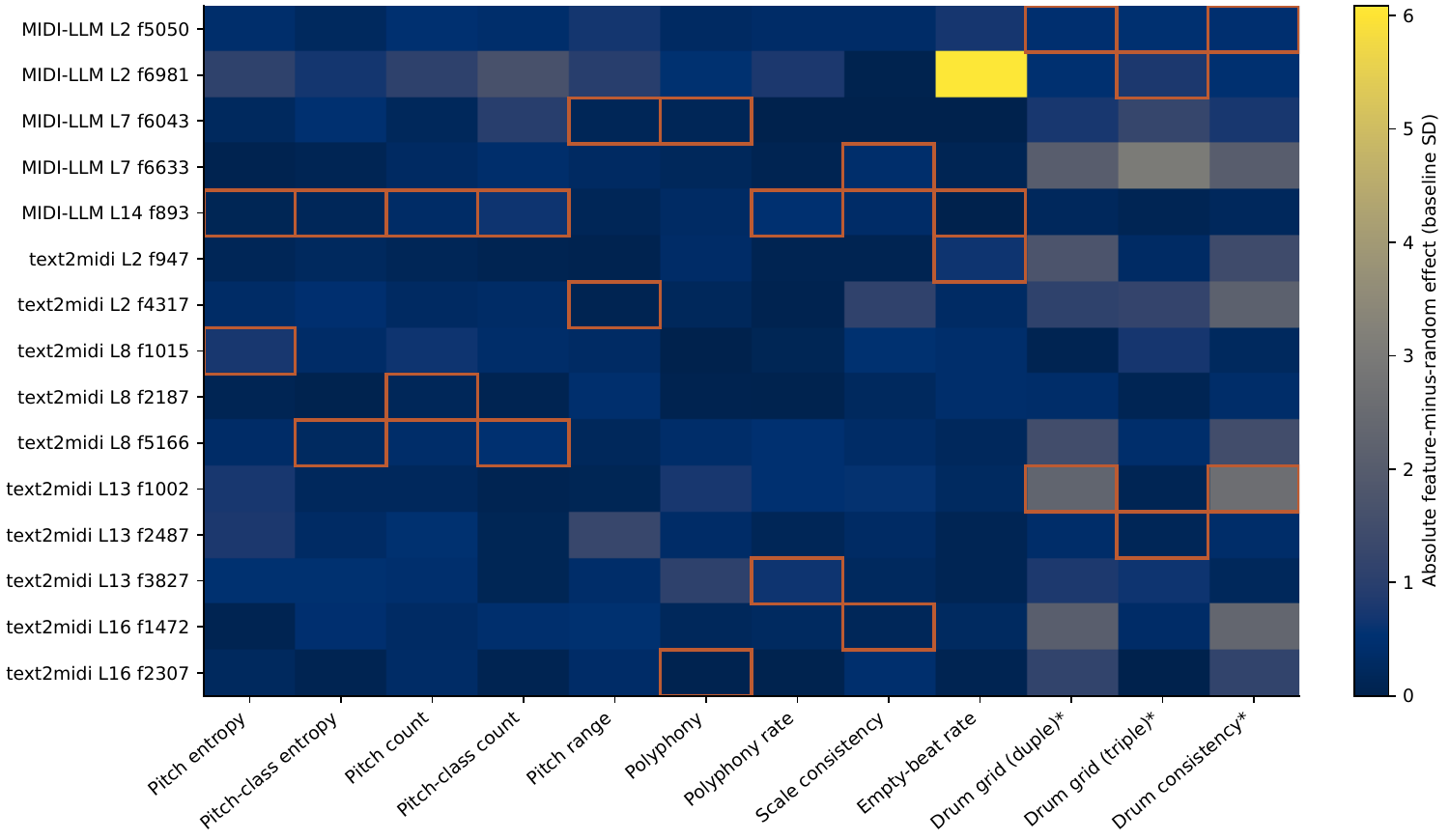}
\caption{Effects of the 15 metric features on all twelve metrics at $P_{90}$. Color:
absolute feature-minus-random effect in baseline SD; outline: hypothesized
metric; asterisk: drum metric, prompts with defined values only.}
\label{fig:cross_metric}
\end{figure}

Tables~\ref{tab:all_muspy} and \ref{tab:all_instruments} list every primary
metric and program test of the targeted screen at $P_{90}$ with its archived
$p$ and $q$ values; the replication for the synth-lead feature (f3340), which
entered the screen as a
positive control, is in Table~\ref{tab:f3340}. Table~\ref{tab:muspy_selected}
gives both references for the effects of Section~\ref{sec:two_references},
Figure~\ref{fig:paired_prompts} shows the paired prompts behind selected
effects, and Figure~\ref{fig:cross_metric} shows the effect of every metric
feature on all twelve metrics. Table~\ref{tab:sensitivity} gives the
resolution of the exact tests and the smallest observed corrected $q$ values.
Benjamini--Hochberg uses the ordered thresholds $k\,0.05/m$ for $m$ hypotheses;
$0.05/m$ alone does not determine whether a hypothesis can pass. None of the
three exploratory screens yielded a primary per-feature rejection at $q<0.05$.

\begin{table}[htbp]
\centering\small\setlength{\tabcolsep}{4pt}
\caption{Exact-test resolution and observed Benjamini--Hochberg results.
The smallest possible $p$ is $2^{-n}$ for the maximum number $n$ of prompts;
undefined pairs or ties can raise this bound. The final column reports the
smallest observed $q$ in each primary family.}
\label{tab:sensitivity}
\begin{tabular}{lrrrr}
\toprule
Screen & Hypotheses & Max. prompts & Smallest possible $p$ & Smallest observed $q$\\
\midrule
Program presence & 27 & 10 & 0.000977 & 1.000 \\
Raw-correlation metric & 24 & 10 & 0.000977 & 0.727 \\
Density-adjusted metric & 42 & 8 & 0.003906 & 0.246 \\
\bottomrule
\end{tabular}
\end{table}

\paragraph{Family-level sensitivity.}
The main intervals resample shared prompts with fixed observed baseline SDs.
All 20,000 draws retain every hypothesis in the density-adjusted family.
For raw correlations, draws with no complete pair for at least one hypothesis
are excluded in full: 19,859, 19,383, and 17,809 draws remain at $P_{50}$,
$P_{75}$, and $P_{90}$, respectively, for each contrast. These intervals
condition on the whole family being evaluable.
Re-estimating SDs within draws produces undefined effects when a resampled
baseline is constant or fewer than two complete pairs remain. We exclude such
draws in full, without dropping individual hypotheses. At $P_{90}$, the
feature-minus-random interval retains 19,937 draws for the density-adjusted
family but only 4,849 for the raw-correlation family. The latter conditional
interval is sensitive to the exclusion of 75.8\% of draws and is not used for
the main inference. For the density-adjusted family at $P_{90}$, the
feature-minus-random interval includes zero under both alternative schemes. Table~\ref{tab:family_sensitivity} also gives the earlier
dictionary-bootstrap intervals, which condition on the observed prompts.

\begin{table}[htbp]
\centering\footnotesize\setlength{\tabcolsep}{3pt}
\caption{Sensitivity of family-level feature-minus-random effects, in baseline
SD. Re-estimated scale: mean effect and 95\% prompt-bootstrap interval after
excluding undefined draws; retained: usable draws out of 20,000; dictionary:
95\% interval resampling dictionaries conditional on the observed prompts.}
\label{tab:family_sensitivity}
\begin{tabular}{llrrr}
\toprule
Family & Target & Re-estimated scale & Retained & Dictionary\\
\midrule
Density-adjusted (42, 21 SAEs) & $P_{50}$ & $+0.02$ $[-0.10, +0.19]$ & 19,945/20,000 & $+0.02$ $[-0.13, +0.18]$ \\
Density-adjusted (42, 21 SAEs) & $P_{90}$ & $+0.17$ $[-0.01, +0.46]$ & 19,937/20,000 & $+0.17$ $[-0.01, +0.34]$ \\
Raw correlation (24, 7 SAEs) & $P_{50}$ & $-0.04$ $[-0.27, +0.36]$ & 12,117/20,000 & $-0.04$ $[-0.13, +0.09]$ \\
Raw correlation (24, 7 SAEs) & $P_{75}$ & $+0.15$ $[-1.00, +0.84]$ & 10,991/20,000 & $+0.15$ $[-0.19, +0.69]$ \\
Raw correlation (24, 7 SAEs) & $P_{90}$ & $-0.23$ $[-1.09, +0.50]$ & 4,849/20,000 & $-0.23$ $[-0.61, +0.11]$ \\
\bottomrule
\end{tabular}
\end{table}

\begin{table}[htbp]
\centering\small\setlength{\tabcolsep}{4pt}
\caption{All 24 primary metric hypotheses of the targeted screen at $P_{90}$.
Oriented difference: feature minus random, signed by the observational
correlation (positive agrees with the hypothesis), in metric units; pairs
exclude undefined values.}
\label{tab:all_muspy}
\begin{tabular}{lllrrrr}
\toprule
Model & Feature & Metric & Pairs & Oriented difference & $p$ & $q$\\
\midrule
MIDI-LLM & L2 f5050 & Drum grid (duple) & 5 & -0.019 & 0.6250 & 0.948 \\
MIDI-LLM & L2 f5050 & Drum consistency & 5 & -0.019 & 0.6250 & 0.948 \\
MIDI-LLM & L2 f6981 & Drum grid (triple) & 2 & +0.042 & 0.5000 & 0.948 \\
MIDI-LLM & L7 f6043 & Pitch range & 9 & -2.111 & 0.6250 & 0.948 \\
MIDI-LLM & L7 f6043 & Polyphony & 9 & -0.260 & 0.7695 & 0.948 \\
MIDI-LLM & L7 f6633 & Scale consistency & 9 & +0.036 & 0.1309 & 0.785 \\
MIDI-LLM & L14 f893 & Empty-beat rate & 10 & -0.000 & 1.0000 & 1.000 \\
MIDI-LLM & L14 f893 & Pitch-class count & 10 & -1.400 & 0.8438 & 0.948 \\
MIDI-LLM & L14 f893 & Pitch count & 10 & -4.700 & 0.8691 & 0.948 \\
MIDI-LLM & L14 f893 & Pitch-class entropy & 8 & -0.091 & 0.7070 & 0.948 \\
MIDI-LLM & L14 f893 & Pitch entropy & 8 & -0.123 & 0.6836 & 0.948 \\
MIDI-LLM & L14 f893 & Polyphony rate & 10 & -0.164 & 0.9766 & 1.000 \\
text2midi & L2 f947 & Empty-beat rate & 10 & -0.025 & 0.5625 & 0.948 \\
text2midi & L2 f4317 & Pitch range & 10 & -0.800 & 0.5625 & 0.948 \\
text2midi & L8 f1015 & Pitch entropy & 10 & +0.418 & 0.0303 & 0.727 \\
text2midi & L8 f2187 & Pitch count & 10 & -1.500 & 0.8535 & 0.948 \\
text2midi & L8 f5166 & Pitch-class count & 10 & +1.100 & 0.0938 & 0.750 \\
text2midi & L8 f5166 & Pitch-class entropy & 10 & +0.114 & 0.1836 & 0.881 \\
text2midi & L13 f1002 & Drum grid (duple) & 6 & -0.114 & 0.7969 & 0.948 \\
text2midi & L13 f1002 & Drum consistency & 6 & -0.129 & 0.8281 & 0.948 \\
text2midi & L13 f2487 & Drum grid (triple) & 6 & -0.029 & 0.7500 & 0.948 \\
text2midi & L13 f3827 & Polyphony rate & 10 & +0.132 & 0.0684 & 0.750 \\
text2midi & L16 f1472 & Scale consistency & 10 & +0.012 & 0.3398 & 0.948 \\
text2midi & L16 f2307 & Polyphony & 10 & +0.083 & 0.4180 & 0.948 \\
\bottomrule
\end{tabular}
\end{table}

\begin{table}[htbp]
\centering\small
\caption{All 27 program hypotheses of the targeted screen beyond the synth-lead feature. Zero-based
programs; counts: program-positive outputs among ten prompts at $P_{90}$.}
\label{tab:all_instruments}
\begin{tabular}{llrrrrrr}
\toprule
Model & Feature & Program & Baseline & Feature & Random & $p$ & $q$\\
\midrule
MIDI-LLM & L2 f920 & 28 & 0 & 0 & 0 & 1.0000 & 1.000 \\
MIDI-LLM & L2 f5165 & 49 & 0 & 0 & 0 & 1.0000 & 1.000 \\
MIDI-LLM & L7 f499 & 0 & 8 & 7 & 8 & 1.0000 & 1.000 \\
MIDI-LLM & L7 f6424 & 24 & 0 & 2 & 0 & 0.2500 & 1.000 \\
text2midi & L2 f571 & 1 & 7 & 7 & 6 & 0.5000 & 1.000 \\
text2midi & L2 f896 & 30 & 2 & 2 & 0 & 0.2500 & 1.000 \\
text2midi & L2 f1409 & 27 & 1 & 1 & 1 & 1.0000 & 1.000 \\
text2midi & L2 f1740 & 28 & 1 & 1 & 2 & 1.0000 & 1.000 \\
text2midi & L2 f2195 & 2 & 2 & 1 & 2 & 1.0000 & 1.000 \\
text2midi & L2 f2686 & 4 & 1 & 2 & 1 & 0.5000 & 1.000 \\
text2midi & L2 f2699 & 51 & 0 & 1 & 0 & 0.5000 & 1.000 \\
text2midi & L2 f3093 & 50 & 1 & 0 & 1 & 1.0000 & 1.000 \\
text2midi & L2 f3905 & 71 & 0 & 0 & 0 & 1.0000 & 1.000 \\
text2midi & L2 f4003 & 34 & 2 & 1 & 2 & 1.0000 & 1.000 \\
text2midi & L2 f4644 & 26 & 0 & 0 & 0 & 1.0000 & 1.000 \\
text2midi & L2 f5962 & 48 & 2 & 2 & 3 & 0.8750 & 1.000 \\
text2midi & L8 f558 & 47 & 0 & 2 & 0 & 0.2500 & 1.000 \\
text2midi & L8 f1354 & 30 & 2 & 1 & 2 & 1.0000 & 1.000 \\
text2midi & L8 f3214 & 52 & 0 & 1 & 0 & 0.5000 & 1.000 \\
text2midi & L8 f3802 & 18 & 0 & 1 & 0 & 0.5000 & 1.000 \\
text2midi & L8 f4126 & 48 & 2 & 2 & 4 & 0.9375 & 1.000 \\
text2midi & L8 f4586 & 29 & 1 & 1 & 1 & 0.7500 & 1.000 \\
text2midi & L8 f4808 & 29 & 1 & 1 & 2 & 1.0000 & 1.000 \\
text2midi & L8 f5932 & 27 & 1 & 2 & 0 & 0.2500 & 1.000 \\
text2midi & L16 f995 & 4 & 1 & 5 & 1 & 0.0625 & 1.000 \\
text2midi & L16 f2977 & 1 & 7 & 5 & 7 & 1.0000 & 1.000 \\
text2midi & L16 f3842 & 5 & 0 & 1 & 1 & 0.7500 & 1.000 \\
\bottomrule
\end{tabular}
\end{table}

\begin{table}[htbp]
\centering\footnotesize\setlength{\tabcolsep}{3pt}
\caption{\tmodel{} metric effects that differ between the two references ($P_{90}$).
Raw metric differences, paired-prompt bootstrap 95\% intervals over ten
prompts; all three hypotheses predict an increase; $p$, $q$: feature minus
random within the 24 metric hypotheses.}
\label{tab:muspy_selected}
\begin{tabular}{llrrrr}
\toprule
Feature & Metric & Feature $-$ baseline & Feature $-$ random & $p$ & $q$\\
\midrule
L8 f1015 & Pitch entropy & 0.271 [-0.102, 0.661] & 0.418 [0.078, 0.783] & 0.0303 & 0.727 \\
L13 f3827 & Polyphony rate & -0.003 [-0.123, 0.112] & 0.132 [-0.004, 0.282] & 0.0684 & 0.750 \\
L8 f5166 & Pitch-class count & 0.100 [-1.000, 1.200] & 1.100 [0.000, 2.400] & 0.0938 & 0.750 \\
\bottomrule
\end{tabular}
\end{table}

\paragraph{Density-adjusted screen.}
\label{app:robustness}
Table~\ref{tab:robustness} lists all 42 features of the density-adjusted
screen. Thirty comparisons use all eight prompts, eleven use seven, and one
uses six, because some outputs have undefined metric values; the first
\mmodel{} baseline prompt produces drums only.

\begin{table}[htbp]
\centering\scriptsize\setlength{\tabcolsep}{3pt}
\caption{All 42 features of the density-adjusted screen at $P_{90}$. Oriented
difference: feature minus random, signed by the adjusted correlation, in metric
units, with paired bootstrap 95\% interval; $q$: Benjamini--Hochberg across the
42 features; final: dictionaries of Table~\ref{tab:health}; historical: earlier
checkpoints (Appendix~\ref{app:selection}).}
\label{tab:robustness}
\begin{tabular}{llllrrrr}
\toprule
Model & Dictionary & Feature & Metric & Pairs & Oriented difference [95\% CI] & $p$ & $q$\\
\midrule
text2midi & L2 EF4 k32 (historical) & f1247 & Pitch range & 8 & $+0.12$ $[-9.00, +9.50]$ & 0.5000 & 0.778 \\
text2midi & L2 EF4 k32 (historical) & f1596 & Polyphony & 8 & $-0.81$ $[-1.56, -0.01]$ & 0.9453 & 0.984 \\
text2midi & L2 EF8 k32 (historical) & f3526 & Pitch range & 8 & $+0.38$ $[-15.00, +11.38]$ & 0.4922 & 0.778 \\
text2midi & L2 EF8 k32 (historical) & f5056 & Polyphony & 8 & $+1.63$ $[+0.68, +2.58]$ & 0.0117 & 0.246 \\
text2midi & L2 EF8 k32 (final) & f1712 & Pitch range & 8 & $+0.62$ $[-21.90, +17.63]$ & 0.4844 & 0.778 \\
text2midi & L2 EF8 k32 (final) & f2208 & Polyphony & 8 & $-0.16$ $[-2.20, +1.68]$ & 0.5469 & 0.792 \\
text2midi & L8 EF8 k32 (historical) & f4666 & Pitch range & 8 & $+1.62$ $[-3.38, +6.00]$ & 0.2852 & 0.778 \\
text2midi & L8 EF8 k32 (historical) & f5379 & Polyphony & 8 & $-0.65$ $[-1.52, +0.21]$ & 0.8945 & 0.963 \\
text2midi & L8 EF8 k32 (final) & f1053 & Pitch range & 8 & $-8.38$ $[-16.38, +0.25]$ & 0.9609 & 0.984 \\
text2midi & L8 EF8 k32 (final) & f3252 & Polyphony & 8 & $+0.25$ $[-0.89, +1.48]$ & 0.3633 & 0.778 \\
text2midi & L13 EF4 k32 (historical) & f10 & Polyphony & 8 & $+0.96$ $[-1.05, +3.18]$ & 0.2461 & 0.778 \\
text2midi & L13 EF4 k32 (historical) & f1869 & Pitch range & 8 & $-3.00$ $[-7.50, +1.38]$ & 0.8828 & 0.963 \\
text2midi & L13 EF8 k32 (historical) & f4933 & Pitch range & 8 & $+1.12$ $[-7.00, +12.75]$ & 0.4648 & 0.778 \\
text2midi & L13 EF8 k32 (historical) & f6039 & Polyphony & 8 & $+0.20$ $[-0.52, +0.95]$ & 0.3125 & 0.778 \\
text2midi & L13 EF8 k32 (final) & f296 & Polyphony & 8 & $+1.17$ $[-0.32, +2.63]$ & 0.1172 & 0.656 \\
text2midi & L13 EF8 k32 (final) & f1608 & Pitch range & 7 & $+15.00$ $[-0.15, +34.71]$ & 0.0703 & 0.591 \\
text2midi & L16 EF8 k32 (historical) & f441 & Polyphony & 8 & $+1.01$ $[-0.38, +2.35]$ & 0.1133 & 0.656 \\
text2midi & L16 EF8 k32 (historical) & f4343 & Pitch range & 8 & $+4.12$ $[-5.00, +11.25]$ & 0.1875 & 0.656 \\
text2midi & L16 EF8 k32 (final) & f858 & Pitch range & 8 & $-3.12$ $[-10.88, +3.88]$ & 0.7734 & 0.963 \\
text2midi & L16 EF8 k32 (final) & f2160 & Polyphony & 8 & $-0.71$ $[-2.00, +0.33]$ & 0.8320 & 0.963 \\
MIDI-LLM & L2 EF4 k32 (historical) & f2012 & Polyphony & 7 & $+0.43$ $[-0.38, +1.17]$ & 0.1797 & 0.656 \\
MIDI-LLM & L2 EF4 k32 (historical) & f8044 & Polyphony rate & 8 & $+0.03$ $[-0.18, +0.25]$ & 0.4297 & 0.778 \\
MIDI-LLM & L2 EF4 k32 (final) & f6774 & Polyphony & 7 & $-0.60$ $[-1.47, +0.41]$ & 0.8750 & 0.963 \\
MIDI-LLM & L2 EF4 k32 (final) & f7370 & Polyphony rate & 8 & $-0.22$ $[-0.40, -0.06]$ & 0.9922 & 0.992 \\
MIDI-LLM & L7 EF4 k32 (historical) & f253 & Polyphony & 7 & $+0.58$ $[-0.43, +1.47]$ & 0.1484 & 0.656 \\
MIDI-LLM & L7 EF4 k32 (historical) & f773 & Polyphony rate & 8 & $+0.01$ $[-0.29, +0.29]$ & 0.4609 & 0.778 \\
MIDI-LLM & L7 EF4 k32 (final) & f4491 & Polyphony rate & 8 & $+0.36$ $[+0.15, +0.57]$ & 0.0078 & 0.246 \\
MIDI-LLM & L7 EF4 k32 (final) & f6043 & Polyphony & 7 & $-0.48$ $[-1.65, +0.91]$ & 0.7578 & 0.963 \\
MIDI-LLM & L12 EF32 k128 (historical) & f54452 & Polyphony & 7 & $-0.21$ $[-0.50, +0.18]$ & 0.8672 & 0.963 \\
MIDI-LLM & L12 EF32 k128 (historical) & f57761 & Polyphony rate & 8 & $+0.15$ $[-0.15, +0.44]$ & 0.1875 & 0.656 \\
MIDI-LLM & L12 EF32 k32 (historical) & f127 & Polyphony rate & 8 & $-0.01$ $[-0.21, +0.23]$ & 0.5234 & 0.785 \\
MIDI-LLM & L12 EF32 k32 (historical) & f8764 & Polyphony & 7 & $+1.31$ $[+0.33, +2.60]$ & 0.0234 & 0.328 \\
MIDI-LLM & L12 EF4 k32 (historical) & f4015 & Polyphony & 6 & $-0.35$ $[-1.09, +0.31]$ & 0.8125 & 0.963 \\
MIDI-LLM & L12 EF4 k32 (historical) & f4528 & Polyphony rate & 8 & $-0.02$ $[-0.33, +0.24]$ & 0.5742 & 0.804 \\
MIDI-LLM & L12 EF4 k32 (final) & f242 & Polyphony rate & 8 & $+0.07$ $[-0.13, +0.21]$ & 0.2656 & 0.778 \\
MIDI-LLM & L12 EF4 k32 (final) & f1783 & Polyphony & 7 & $+0.94$ $[-0.72, +2.52]$ & 0.1484 & 0.656 \\
MIDI-LLM & L14 EF32 k128 (historical) & f20313 & Polyphony & 7 & $+0.11$ $[-0.91, +1.24]$ & 0.4453 & 0.778 \\
MIDI-LLM & L14 EF32 k128 (historical) & f63199 & Polyphony rate & 8 & $+0.03$ $[-0.14, +0.16]$ & 0.3594 & 0.778 \\
MIDI-LLM & L14 EF4 k32 (historical) & f1215 & Polyphony & 7 & $+0.05$ $[-1.03, +1.41]$ & 0.4688 & 0.778 \\
MIDI-LLM & L14 EF4 k32 (historical) & f4433 & Polyphony rate & 8 & $+0.03$ $[-0.22, +0.28]$ & 0.4141 & 0.778 \\
MIDI-LLM & L14 EF4 k32 (final) & f5420 & Polyphony rate & 8 & $+0.26$ $[+0.07, +0.51]$ & 0.0469 & 0.492 \\
MIDI-LLM & L14 EF4 k32 (final) & f5640 & Polyphony & 7 & $-0.20$ $[-1.18, +0.88]$ & 0.6406 & 0.868 \\
\bottomrule
\end{tabular}
\end{table}

\paragraph{Diagnostics.}
\label{app:diagnostics}
Table~\ref{tab:diagnostics} averages the diagnostics over the ten prompts of
each feature in the targeted screen. Baseline generations are shared within
each model, with mean note counts of 105.6 for \tmodel{} and 255.6 for
\mmodel{}, whose count is constrained by its token budget and event
representation. The random branch computes its scalar from its own trajectory,
so realized norms can differ from the feature branch once the trajectories
diverge.

\begin{table}[htbp]
\centering\small\setlength{\tabcolsep}{5pt}
\caption{Intervention diagnostics of the targeted screen. Mean $\pm$ SD across 52
\tmodel{} or 28 \mmodel{} feature means; relative norm: $100\|h'-h\|_2/\|h\|_2$;
firing: percentage of edited states whose target coordinate is positive after
re-encoding; induction at $P_{90}$; ablation removes the naturally active
contribution.}
\label{tab:diagnostics}
\begin{tabular}{llrrr}
\toprule
Model & Condition & Notes & Relative norm (\%) & Re-encoded firing (\%)\\
\midrule
text2midi & Feature & 107.4 $\pm$ 5.8 & 21.4 $\pm$ 16.2 & 55.84 $\pm$ 42.88 \\
text2midi & Random & 108.7 $\pm$ 5.6 & 21.4 $\pm$ 16.3 & 0.29 $\pm$ 0.60 \\
text2midi & Ablation & 105.4 $\pm$ 1.1 & 0.0 $\pm$ 0.1 & 0.02 $\pm$ 0.05 \\
MIDI-LLM & Feature & 255.4 $\pm$ 0.5 & 12.9 $\pm$ 10.8 & 35.89 $\pm$ 40.07 \\
MIDI-LLM & Random & 255.5 $\pm$ 0.4 & 13.4 $\pm$ 12.3 & 0.07 $\pm$ 0.18 \\
MIDI-LLM & Ablation & 255.6 $\pm$ 0.0 & 0.0 $\pm$ 0.0 & 0.00 $\pm$ 0.01 \\
\bottomrule
\end{tabular}
\end{table}
\FloatBarrier

\section{Feature-Level Blinded LLM Ratings}
\label{app:semantic_features}

\begin{figure}[!htb]
\centering
\includegraphics[width=\linewidth]{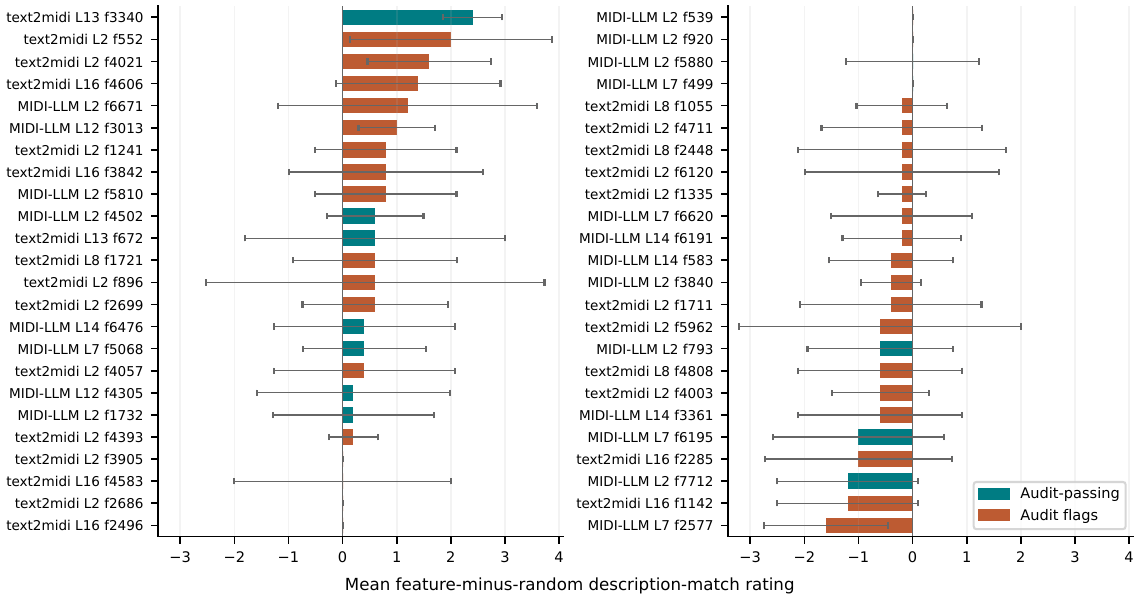}
\caption{Rating differences for all 48 description candidates. Bars: mean over five
paired prompts at $P_{90}$ on the 1--5 scale; error bars: SD; color: audit
status, which defines the primary subset.}
\label{fig:semantic_all}
\end{figure}

Table~\ref{tab:semantic} gives the pooled rating tests,
Table~\ref{tab:semantic_examples} lists the six largest feature-level rating
differences, ordered by the observed difference, and
Figure~\ref{fig:semantic_all} shows all 48.

\begin{table}[htbp]
\centering\small\setlength{\tabcolsep}{4pt}
\caption{Pooled blinded LLM ratings show no advantage of the feature direction.
Rating differences: averaged over five prompts per feature, then over
features; intervals resample features; correct choices: decisions that pick a
variant; last row: primary subset without the synth-lead feature.}
\label{tab:semantic}
\begin{tabular}{lrrrr}
\toprule
Subset & Features & Correct choices & Rating difference [95\% CI] & $p$\\
\midrule
Audit-passing subset & 12 & 4/8 & 0.167 [-0.300, 0.700] & 0.3125 \\
All descriptions & 48 & 16/34 & 0.108 [-0.112, 0.338] & 0.1885 \\
Audit-passing, without synth lead & 11 & 3/7 & -0.036 [-0.400, 0.291] & 0.6133 \\
\bottomrule
\end{tabular}
\end{table}

\begin{table}[htbp]
\centering\footnotesize\setlength{\tabcolsep}{3pt}
\caption{The six largest feature-level rating differences. $\Delta$: mean
feature-minus-random rating, SD over five prompts; favored: prompts with the
feature output rated higher; flags: token concentration (T), token syntax (S),
note-count imbalance (N).}
\label{tab:semantic_examples}
\begin{tabularx}{\linewidth}{l>{\raggedright\arraybackslash}Xrcc}
\toprule
Feature & Description & $\Delta\pm\mathrm{SD}$ & Favored & Flags\\
\midrule
text2midi L13 f3340 & Synth-lead (Lead 2 sawtooth) / electronic-patch instrumentation & $+2.40 \pm 0.55$ & 5/5 & None \\
text2midi L2 f552 & Dense, near-continuous drum/percussion texture & $+2.00 \pm 1.87$ & 4/5 & T, S \\
text2midi L2 f4021 & Rich, multi-timbral instrumentation (many melodic instruments) & $+1.60 \pm 1.14$ & 4/5 & T, S \\
text2midi L16 f4606 & Sparse texture with recurring timpani-type low percussion & $+1.40 \pm 1.52$ & 3/5 & S \\
MIDI-LLM L2 f6671 & Prominent, continuous drum-kit layer relative to melodic notes & $+1.20 \pm 2.39$ & 3/5 & T, S, N \\
MIDI-LLM L12 f3013 & Slow harmonic rhythm with long sustained notes over a static backdrop & $+1.00 \pm 0.71$ & 4/5 & N \\
\bottomrule
\end{tabularx}
\end{table}

\FloatBarrier